\documentclass[twocolumn]{aastex63}
\usepackage[]{amsmath}
\usepackage[]{amssymb}

\newcommand{\hst}{{\it HST}}
\newcommand{\Mgtwo}{Mg~{\sc ii}~}
\newcommand{\Fetwo}{Fe~{\sc ii}~}
\newcommand{\wmgone}{$W_r(\lambda$2796)~}
\newcommand{\wmgtwo}{$W_r(\lambda$2803)~}
\newcommand{\wfeone}{$W_r(\lambda$2600)~}
\newcommand{\w}{$W_r$~}
\newcommand{\kms}{km s$^{-1}$}
\newcommand{\z}{$z$}
\received{}
\revised{}
\accepted{}
\submitjournal{ApJ}

\shorttitle{\Mgtwo and \Fetwo in GNIRS}
\shortauthors{Zou et al.}

\begin{document}

\title{Strong Mg~{\sc ii} and Fe~{\sc ii} Absorbers at 2.2 $<~z~<$ 6.0}

\author[0000-0002-3983-6484]{Siwei Zou}
\affiliation{Kavli Institute for Astronomy and Astrophysics, Peking University, Beijing 100871, China}

\author[0000-0003-4176-6486]{Linhua Jiang}
\affiliation{Kavli Institute for Astronomy and Astrophysics, Peking University, Beijing 100871, China}
\affiliation{Department of Astronomy, School of Physics, Peking University, Beijing 100871, China}

\author[0000-0003-1659-7035]{Yue Shen}
\affiliation{Department of Astronomy, University of Illinois at UrbanaChampaign, Urbana, IL 61801, USA}
\affiliation{National Center for Supercomputing Applications, University of Illinois at Urbana-Champaign, Urbana, IL 61801, USA}

\author[0000-0001-5364-8941]{Jin Wu}
\affiliation{Kavli Institute for Astronomy and Astrophysics, Peking University, Beijing 100871, China}
\affiliation{Department of Astronomy, School of Physics, Peking University, Beijing 100871, China}

\author[0000-0002-2931-7824]{Eduardo Ba{\~n}ados}
\affiliation{Max Planck Institut f\"{u}r Astronomie, K\"{o}nigstuhl 17, D-69117, Heidelberg, Germany}

\author[0000-0003-3310-0131]{Xiaohui Fan}
\affiliation{Steward Observatory, University of Arizona, 933 N Cherry Avenue, Tucson, AZ 85721, USA}

\author[0000-0001-6947-5846]{Luis C. Ho}
\affiliation{Kavli Institute for Astronomy and Astrophysics, Peking University, Beijing 100871, China}
\affiliation{Department of Astronomy, School of Physics, Peking University, Beijing 100871, China}

\author[0000-0001-9585-1462]{Dominik A. Riechers}
\affiliation{Cornell University, Space Sciences Building, Ithaca, NY 14853, USA}

\author[0000-0001-9024-8322]{Bram Venemans}
\affiliation{Max Planck Institut f\"{u}r Astronomie, K\"{o}nigstuhl 17, D-69117, Heidelberg, Germany}

\author[0000-0001-9191-9837]{Marianne Vestergaard}
\affiliation{Steward Observatory, University of Arizona, 933 N Cherry Avenue, Tucson, AZ 85721, USA}
\affiliation{Niels Bohr Institute, University of Copenhagen, Jagtvej 128, DK-2200 Copenhagen, Denmark}

\author[0000-0003-4793-7880]{Fabian Walter}
\affiliation{Max Planck Institut f\"{u}r Astronomie, K\"{o}nigstuhl 17, D-69117, Heidelberg, Germany}

\author[0000-0002-7633-431X]{Feige Wang}
\affiliation{Steward Observatory, University of Arizona, 933 N Cherry Avenue, Tucson, AZ 85721, USA}

\author[0000-0002-4201-7367]{Chris J. Willott}
\affiliation{NRC Herzberg, 5071 West Saanich Road, Victoria, BC V9E 2E7, Canada}

\author[0000-0002-5535-4186]{Ravi Joshi}
\affiliation{Kavli Institute for Astronomy and Astrophysics, Peking University, Beijing 100871, China}

\author[0000-0002-7350-6913]{Xue-Bing Wu}
\affiliation{Kavli Institute for Astronomy and Astrophysics, Peking University, Beijing 100871, China}
\affiliation{Department of Astronomy, School of Physics, Peking University, Beijing 100871, China}

\author[0000-0001-5287-4242]{Jinyi Yang}
\affiliation{Steward Observatory, University of Arizona, 933 N Cherry Avenue, Tucson, AZ 85721, USA}
%%%%%%%%
\begin{abstract} 

We present a study of strong intervening absorption systems in the near-IR spectra of 31 luminous quasars at $z>5.7$. The quasar spectra were obtained with {\it Gemini} GNIRS that provide continuous wavelength coverage from $\sim$0.9 to $\sim$2.5 $\mu$m. We detect 32 strong \Mgtwo doublet absorbers with rest-frame equivalent width \wmgone $>1.0$ \AA\ at $2.2 <z< 6.0$. Each \Mgtwo absorber is confirmed by at least two associated \Fetwo absorption lines in the rest-frame wavelength range of $\sim 1600-2600$ \AA. We find that the comoving line density ($dN/dX$) of the strong Fe~{\sc ii}-bearing \Mgtwo absorbers decreases towards higher redshift at $z>3$, consistent with previous studies. Compared with strong \Mgtwo absorbers detected in damped Ly$\alpha$ systems at 2 $<z<$ 4, our absorbers are potentially less saturated and show much larger rest-frame velocity widths. This suggests that the gas traced by our absorbers are potentially affected by galactic superwinds. We analyze the {\it Hubble Space Telescope} near-IR images of the quasars and identify possible associated galaxies for our strong absorbers. There are a maximum two galaxy candidates found within 5$\arcsec$ radius of each absorber. The median F105W-band magnitude of these galaxy candidates is 24.8 mag, which is fainter than the $L^*$ galaxy luminosity at $z\sim$ 4.  By using our observed $dN/dX$ of strong Mg~{\sc ii} absorbers and galaxy candidates median luminosity, we suggest that at high redshift, strong Mg~{\sc ii} absorbers tend to have a more disturbed environment but smaller halo size than that at $z <$ 1.%We did not detect strong C~{\sc iv} absorption lines (at $z\ge5$) in our spectra. The upper limit of the comoving line density indicates that either C~{\sc iv} lines associated with strong \Mgtwo absorbers are very weak or the number density of strong C~{\sc iv} absorbers decreases rapidly at $z\ge5$.
\end{abstract}


%%%%%%%%%%%%%%%%%%%
\section{Introduction}\label{sec_intro}

The Circumgalactic medium (CGM) is defined as the gas around the disk or interstellar medium of a galaxy typically within the virial radius of the galaxy. Previous studies suggested that the physical conditions of the gas in the CGM are influenced by both cold accretion inflows and galactic outflows (see \citet{tum17} for a review and references therein). Studies of absorption lines towards bright background sources such as quasars provide a unique and powerful tool to study the physical conditions of the gas. Among these absorption lines, the low-ionization \Mgtwo $\lambda\lambda$2796,2803 doublet is found to be associated with cool components (T$\sim$10$^4$ K) in CGM \citep{ber91,ste02}. Observationally, the connection between the \Mgtwo absorption and CGM is studied using quasar-galaxy pairs at low redshift. By comparing the kinematics of absorbers and galaxies, \Mgtwo has been shown to trace both metal-enriched infalling gas \citep{chen10,lov11,kac11,rub12,bou13a,zab19}, and outflows from luminous star-forming galaxies \citep{bou06,mar09,not10,men12,sch16,sch19}. \citet{zab19} studied 9 quasar-galaxy pairs that were selected from 79 \Mgtwo absorbers at $z\sim$1. They found that the halo gas probed by \Mgtwo lines is approximately aligned with the galaxy's angular momentum vector, which suggests that the \Mgtwo gas co-rotates with galaxy disks. Using the same catalog of \Mgtwo absorbers, \citet{sch19} selected 26 quasar-galaxy pairs and studied their azimuthal angle, which is the angle between the galaxy's major axis and quasar location (see e.g. \citealt{zab19} Figure 1). The bimodality of azimuthal angles suggests that the outflows are bi-conical in nature. 

Strong \Mgtwo systems, defined by their rest-frame equivalent width $W_r$, are found to trace cosmic star formation rate (SFR) \citep{men11}.  Observations have shown that \Mgtwo absorbers are associated with a large amount of neutral gas \citep{lan87,ste95,ste97,rao06,nes07}. \citet{rao06} studied 197 \Mgtwo systems and their {H~\sc i} profiles at 0.11 $< z <$ 1.65 using $Hubble~ Space ~Telescope$ ($HST$) UV spectroscopy. Their results show that all the damped Ly$\alpha$ (DLA) systems ( log N({H~\sc i}) [cm$^{-2}]>$ 20.3) have $W_r$ ($\lambda$2796) $>$ 0.6 \AA. As DLA systems are regarded as the progenitors of star-forming galaxies today, consequently, strong \Mgtwo absorption systems are thought to be correlated with star formation as well. As pointed out in \citet{mat12},  systems traced by strong \Mgtwo absorbers tend to belong to galaxies with high SFRs. In the literature, some studies define systems with \wmgone $>$ 0.3 \AA\ as strong systems, while others use \wmgone $> 1.0$ \AA. In this paper, we use the latter as the definition of strong \Mgtwo absorbers.

To trace star formation with strong \Mgtwo absorption systems, the first parameter to calculate is the pathlength number density. The pathlength can be redshift ($dz$) or co-moving pathlength ($dX$), for which
\begin{equation}
    X(z) = \int^z_0(1+z')^2\frac{H_0}{H(z')}dz.
\end{equation}
The number of absorbers per unit redshift (per absorption distance) $dN/dz$ ($dN/dX$) has been studied in strong \Mgtwo systems at low to high redshift. The evolution of the co-moving line density $dN/dX$ is above and beyond passive evoultion due to the expansion of the Universe. At $z<2$, \citet{zhu13a} found that the $dN/dz$ of \Mgtwo absorbers rises with increasing redshift. At $z>2$, \cite{mat12} and \citet{sfs17} (hereafter M12 and C17 respectively) show that the comoving $dN/dz$ of strong \Mgtwo ($W_r > $ 1 \AA) decreases with increasing redshift, by analyzing 110 absorbers at 1.98 $\leq$ $z$ $\leq$ 5.3. In contrast, the comoving $dN/dz$ of weak \Mgtwo systems with \wmgone $<1$ \AA\ is nearly constant over cosmic time \citep{nes05,mat12,sfs17}, which is quite different from that of strong systems.
 
In this paper, we present a sample of strong \Mgtwo absorbers detected in the near-IR spectra of 31 quasars at $z>5.7$ and study the evolution of their number density at $2.2 < z < 6.0$. We also explore possible connections between the absorbers and properties of the associated galaxies. This paper is presented as follows. We introduce our sample and absorption detection method in Section \ref{sec_obs}. The results are presented in Section \ref{sec_stat}. We discuss possible galaxy counterparts in Section \ref{sec_disussion}. Throughout the paper, all magnitudes are expressed in the AB system. The standard cosmology parameters are used: $H_0$ = 70 km s$^{-1}$ Mpc$^{-1}$, $\Omega_\Lambda$ = 0.7 and $\Omega_m$ = 0.3. 

\section{Data and detection of \Mgtwo absorbers}\label{sec_obs}

The quasar near-IR spectra used in this paper were from a large Gemini-GNIRS program \citep{shen19}. Shen et al. observed most of the 52 quasars at $z >$ 5.7 \citep{jiang16} and this program was carried out during 15B-17A semester. By excluding those quasars that already have reasonable good quality spectra, the final sample consists of 50 quasars. Most of these 50 quasars were initially selected from SDSS with a color cut of $i-z > $ 2.2 and have no detection in $ugr$ bands \citep{jiang16}. The observations were executed using the standard ABBA method. A cross-dispersion mode was used to cover the wavelength range from 0.85 to 2.5 $\mu$m.  We use a slit width of 0.675$\arcsec$ that delivers a resolving power R $\sim$ 800 ($\sim$ 376 km s$^{-1}$) (GNIRS mean resolution) with a pixel scale of 0.15\arcsec/pix. The spectral resolution is estimated from the average FWHM of weak and unblended emission lines in the arc file. The emission redshifts of quasars in the sample were measured from a series of  lines (Mg~{\sc ii}, C~{\sc iii}], Si~{\sc iii}, Al~{\sc iii}, C~{\sc iv}, He~{\sc ii}, O~{\sc iii}], Si~{\sc iv}), which takes the velocity shifts of each line into account \citep{shen19}.  The updated redshifts may differ from the original redshifts in the discovery papers, which are with optical spectra only (see Table \ref{table_ew}). The median emission redshift uncertainty is $\sim$ 300 km s$^{-1}$. The GNIRS data were reduced by the combination of two pipelines, PyRAF-based XDGNIRS \citep{mas12}  and the IDL-based XIDL package. The details are described in \citet{shen19}.

We clarify that the quasar colors in our sample are consistent with that at lower redshift, hence, there is very limited bias caused by the background quasars for absorption candidates selection. Color bias of the background quasars in large samples would possibly affect the foreground absorbers selection. For example, in \citet{pro09}, they found an elevated incidence of Lyman limit opacity in the intergalactic medium. This is related to the SDSS quasar selection bias at $z$ = 3.5 to $z = 3.6$. Considering our sample size and quasar colors, this effect, if any, would be within errors and not affect significantly the absorption study results. Also, we did not select absorption candidates based on any presumptions of N (H~{\sc i}). The selection process of absorption candidates is presented in details in Section \ref{sec_alg}.

%---------------------------------------------------------------------------------------------------------------------------------------
\subsection{Detection Algorithm}\label{sec_alg}

We selected 31 quasars with signal-to-noise ratios (S/N) greater than 10. The S/N is a mean S/N per resel measured from the `clean' continuum region of the spectra without strong OH skylines or water vapor absorption features. The mean S/N values of all spectra are presented in Table \ref{table_ew}. Given the low resolution (R$\sim$ 800) of GNIRS spectra, we did not use the lower-S/N spectra. We first fitted each quasar spectrum with a continuum. The continuum was selected interactively with knots in the absorption free wavelength region. The region between two knots was fitted with a spline curve. Then the spectrum was normalized with this continuum. We then used our algorithm to automatically search and identify metal absorbers in the normalized spectra. The absorption feature was identified with a Gaussian kernel filter, which has a rest-frame velocity FWHM between 376 km s$^{-1}$ and 600 km s$^{-1}$  (six pixels, empirically selected). If $W_r$ of this Gaussian kernel is greater than 0.8 \AA , which is our detection limit (e.g. observe equivalent width $\sim$ 3 \AA\ around wavelength 10,000 \AA) for Mg~{\sc ii} line, then it was considered as an absorption feature. The $W_r$ was measured from the flux summation over $\Delta \lambda$ where the Gaussian kernel is within 3\% of the continuum. For Mg~{\sc ii} doublet, the two kernels of the doublet are separated by $\sim$ 770 km s$^{-1}$ and cross-correlated with the spectrum simultaneously. The selection criteria of Mg~{\sc ii} candidates relate to $W_r,\sigma(W_r)$ and S/N in the continuum. We calculated the $\sigma(W_r$) by using a method by \citet{vol06}. For a normalized spectrum, the $W_r$ of an absorption line is defined as:
\begin{equation}\label{eq:1}
W_r = \int_{\lambda_2}^{\lambda_1} (1-F) d\lambda \approx \Delta\lambda_r (1- \overline{F}),\\
\end{equation}
where $\Delta \lambda_r$ = $(\lambda_2-\lambda_1)/(1+z)$ is the rest-frame absorption line width. $\overline{F}$ is the mean normalized flux density of the absorption line. Equation \ref{eq:1} can be expanded in a Taylor series: 
\begin{equation}
W_r = W_r(\overline{F}) +  \frac{\partial W_r}{\partial \overline{F}} \sigma(F),
\end{equation}
According to Equation \ref{eq:1}, there is $\frac{\partial W_r}{\partial \overline{F}} = -\Delta\lambda$. Together with $\sigma(F)$ = $\frac{\overline{F}}{\textrm{S/N}_c}$, we have
\begin{equation}
\sigma(W_r) = \Delta\lambda\times\frac{\overline{F}}{(\mathrm{S/N)}_c}.
\end{equation}
(S/N)$_c$ is the average S/N per resel of $\pm$ 10 pixels adjacent to $\Delta\lambda$. %This method considers uncertainties in both absorption profile and the continuum which results from S/N. 
The specific Mg~{\sc ii} candicates selection criteria are in the following:

1) \wmgone / $\sigma~(W_r)~>$ 3. 

2) \wmgone $>$ 0.8 \AA\ and \wmgtwo $>$ 0.4 \AA, 

3) S/N $>$ 3 per resel in three or more contiguous pixels beyond the $\Delta\lambda$ region.

\begin{figure*}
  \includegraphics[width=190mm]{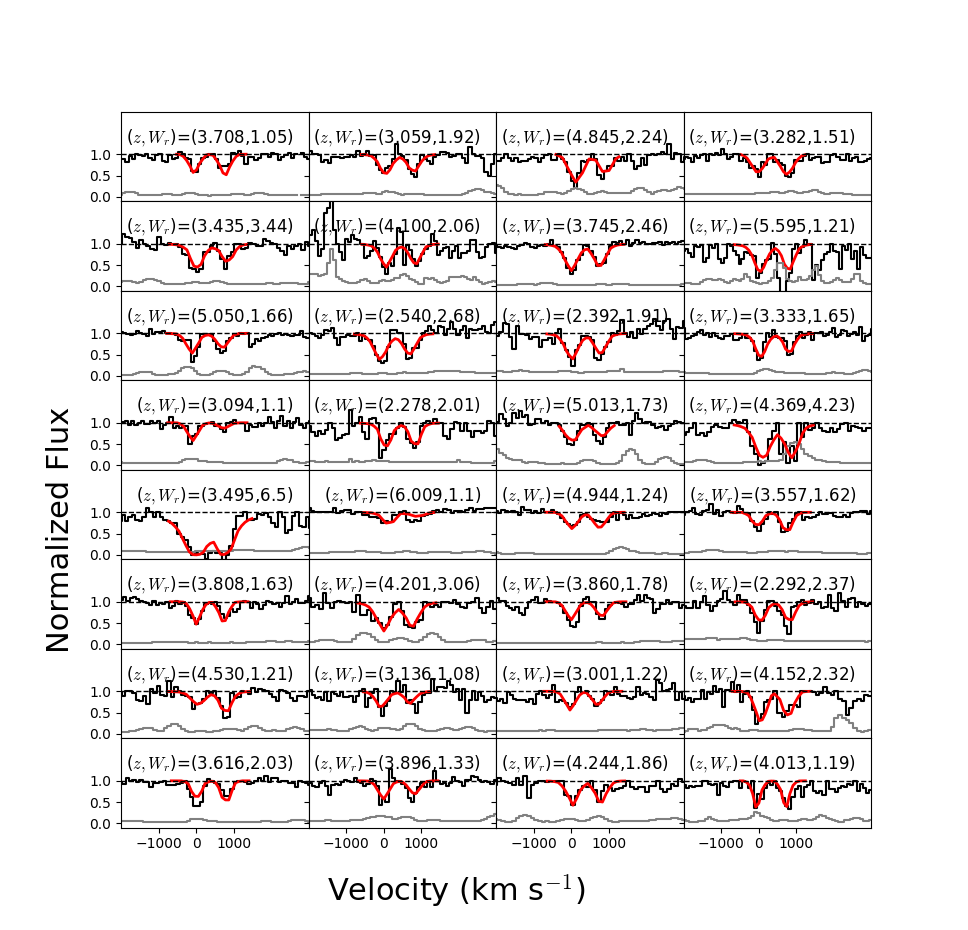}
  \caption{\small{All strong \Mgtwo absorbers in this work. Absorption line redshift $z$ and $W_r$ (\AA) are labeled for each absorber. The median $W_r$ for the all the Mg~{\sc ii} absorbers is 1.78 \AA\ and median absorbers redshfit is $z$ = 3.743. The black and grey curves are the normalized spectra and noise spectra, respectively. The red curves are the best-fitted Voigt profiles. Each absorber is centered at the absorption profile of Mg {\sc ii} $\lambda$2796.
}}\label{fig:wmg_vmg}
\end{figure*}

\begin{figure}
  \resizebox{\hsize}{!}{\includegraphics{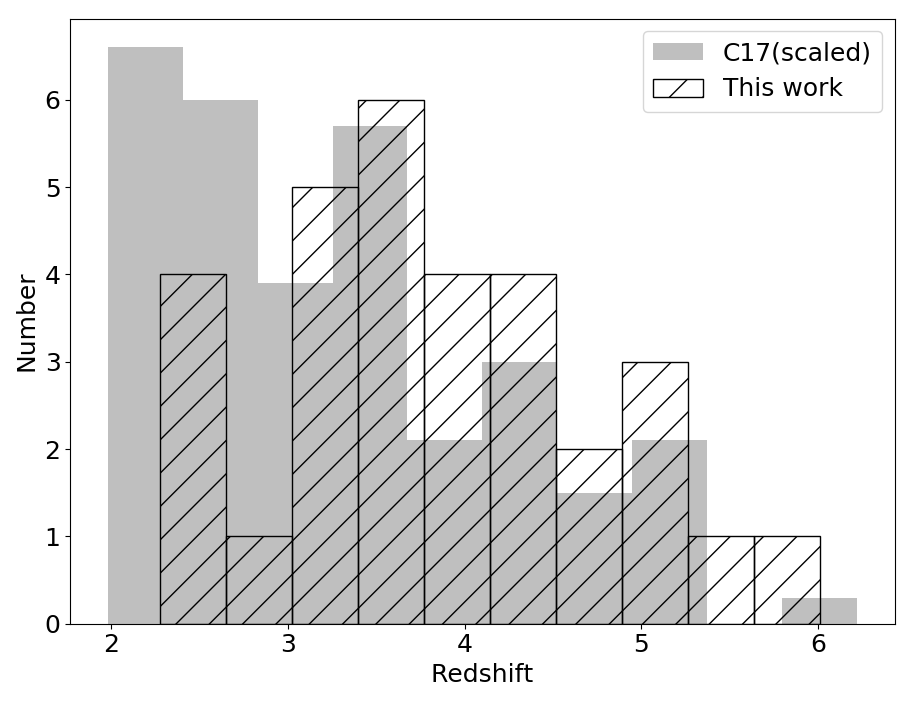}}
  \caption{\small{ Redshift distribution of strong \Mgtwo absorbers ($W_r >$ 1 \AA) in this work (hashed). The redshift distribution of \citet{sfs17} is scaled by a factor of 0.30 for comparison (in gray).
}}\label{fig:z_hist}
\end{figure}

We searched for \Mgtwo systems in all 31 quasar spectra using the above criteria and obtained 110 candidates. Afterward, at least two Fe~{\sc ii} lines (at 1608, 1611, 2344, 2374, 2586, or 2600 \AA) were visually inspected at the same redshift to further confirm the identified \Mgtwo doublet. In the end, we confirmed 32 \Mgtwo and Fe~{\sc ii} absorbers at 2.2 $ <z<$ 6.0. The spectra of the absorbers are presented in Figure \ref{fig:wmg_vmg}. We found that all these \Mgtwo absorbers have \wmgone $>$ 1.0 \AA, and 13 of them are very strong with \wmgone $>$ 2.0 \AA. The median $W_r$ is 1.86 \AA. The redshift distribution (with a median $z$ = 3.743) of these absorbers is shown in Figure \ref{fig:z_hist}. 
% C17 median $z$ = 3.311, respectively. }

\begin{table*}
\begin{center}

  \caption{Summary of 32 strong absorbers.}
\begin{tabular}{lccccccccc}
\hline
(1) Quasar & (2) $z_{em}$ & (3) $z_{abs}$& (4) $W_r$($\lambda$2796)  & (5) $W_r$($\lambda$2803)  & (6) $W_r$($\lambda$2600)   & (7) $\Delta v$  &(8) $\Delta v_\textrm{obs}$ &  (9) S/N\\
 &&& (\AA)  & (\AA)  & (\AA)  &(km s$^{-1}$) &(km s$^{-1}$)&\\
\hline
P000+26      & 5.733  &  3.708  &   1.05$\pm$0.28  &  0.90$\pm$0.23 &   -                        &      $<$152          & $<$435    & 18  \\   
J0002+2550 & 5.818  &  3.059  &   1.92$\pm$0.38  &  1.86$\pm$0.16 &   -                        &      478$\pm$38  & 707$\pm$22    &  18  \\
J0008-0626  & 5.929  &    -       &    -                       &    -                      &    -                        &     -                      &-         & 10    \\
J0028+0457 & 5.982  &  4.845  &   2.24$\pm$0.76  &  1.65$\pm$0.58 &   -                        &      379$\pm$37   &717$\pm$20     &  10  \\
                     &         &  3.282   &   1.51$\pm$0.48  &  1.76$\pm$0.49 &  0.91$\pm$0.43    &      232$\pm$36   &640$\pm$18    &        \\
J0050+3445 & 6.251  &  3.435  &   3.44$\pm$0.88  &  2.02$\pm$0.56 &  1.05$\pm$0.39  &       609$\pm$39   &820$\pm$24    &  10  \\
J0203+0012 & 5.709  &    -       &    -                       &    -                      &    -                        &    -   &-                              &  16  \\
J0300-2232 &  5.684  &  4.100  &   2.06$\pm$0.86  &  1.51$\pm$0.63 &  0.67$\pm$0.32  &     353$\pm$37     &690$\pm$20    &  13  \\
J0353+0104 & 6.057  &    -       &    -                       &    -                      &    -                        &    -     &-                            & 13  \\
J0810+5105 & 5.805  &    -       &    -                       &    -                      &    -                        &    -       &-                           & 13  \\
J0836+0054 & 5.834  &  3.745  &   2.46$\pm$0.44  &  1.84$\pm$0.42 &  0.66$\pm$0.29  &      548$\pm$39   &785$\pm$24    &  20  \\
J0840+5624 & 5.816  &  5.595  &  2.74$\pm$0.25   &  2.57$\pm$0.10 &  0.71$\pm$0.28  &      194$\pm$36   &661$\pm$18   &  17  \\
J0842+1218 & 6.069  &  5.050  &   1.66$\pm$0.57  &  1.25$\pm$0.36 &  2.04$\pm$0.54  &      316$\pm$37   &660$\pm$20   &  13 \\
                    &          &  2.540  &   2.68$\pm$0.51  &  1.75$\pm$0.73 &  0.90$\pm$0.36  &      310$\pm$37     &661$\pm$20    &       \\
                    &          &  2.392  &   2.01$\pm$0.30  &  1.91$\pm$0.24 &  1.21$\pm$0.50  &      334$\pm$37      &709$\pm$20    &       \\
J0850+3246 & 5.730  &  3.333  &   1.65$\pm$0.46  &  1.18$\pm$0.32 &  1.15$\pm$0.33  &      165$\pm$37   &494$\pm$20     &  17 \\
                    &          &  3.094  &   1.10$\pm$0.34  &  0.38$\pm$0.35 &  1.03$\pm$0.38  &      200$\pm$37     &440$\pm$20       &       \\
J0927+2001 & 5.770  &    -       &    -                       &    -                      &    -                        &    -      &-              &  10  \\
J1044-0125 & 5.780  &  2.278  &   2.01$\pm$0.31  &  1.76$\pm$0.32 &  0.71$\pm$0.23  &      155$\pm$34    &398$\pm$14      &  19  \\
J1137+3549 & 6.009  &  5.013  &   1.73$\pm$0.57  &  1.32$\pm$0.45 &  0.96$\pm$0.48  &      481$\pm$38   & 702$\pm$22 &  10  \\
J1148+0702 & 6.344  &  4.369  &   4.23$\pm$0.52  &  2.88$\pm$0.43 &  3.61$\pm$0.54  &     410$\pm$38    & 690$\pm$22 & 12  \\
         %         &          &  3.632  &   1.04$\pm$0.25  &  1.00$\pm$0.22 &  0.59$\pm$0.40  &      276$\pm$35     &  13  \\
                    &          &  3.495\footnote{This Mg~{\sc ii} doublet is strongly blended, so measurements have inevitably large uncertainties.}  &   6.50$\pm$1.20  &  6.3$\pm$0.70 & 1.64$\pm$0.46  &      $>$865 & $>$1059   &  13  \\
J1148+5251 & 6.416  &  6.009  &   1.10$\pm$0.34  &  0.75$\pm$0.27 &  1.63$\pm$0.18  &     207$\pm$38   &611$\pm$22    &  18  \\
                    &          &  4.944  &   1.24$\pm$0.46  &  1.21$\pm$0.42 &  0.34$\pm$0.29  &      $<$103             &$<$613     &       \\
                    &          &  3.557  &   1.62$\pm$0.30  &  1.74$\pm$0.21 &  0.74$\pm$0.25  &      349$\pm$35     &400$\pm$16    &        \\
J1207+0630 & 6.028  &  3.808  &   1.63$\pm$0.40  &  1.50$\pm$0.45 &  1.66$\pm$0.47  &      456$\pm$36   &663$\pm$18    &  10  \\
J1243+2529 & 5.842  &    -       &    -                        &    -                      &    -                       &     -&-                              &   11  \\
J1250+3130 & 6.138  &  4.201  &   3.06$\pm$0.52  &  2.68$\pm$0.51 &  0.73$\pm$0.52  &      179$\pm$38   &530$\pm$22    &  12   \\
                    &          &  3.860  &   1.78$\pm$0.60  &  1.01$\pm$0.49 &  1.80$\pm$0.58  &      297$\pm$35      &630$\pm$16    &         \\
                    &          &  2.292  &   2.37$\pm$0.42  &  1.92$\pm$0.47 &    -                       &      381$\pm$37     & 699$\pm$20   &         \\
J1257+6349 & 5.992  &    -       &    -                       &    -                      &    -                        &    -     &-            &  11  \\    
J1335+3533 & 5.870  &  4.530  &   1.21$\pm$0.45  &  1.27$\pm$0.30 &  1.33$\pm$0.47  &      323$\pm$38   &545$\pm$22   &  15  \\
J1425+3254 & 5.862  &  3.136  &   1.08$\pm$0.39  &  1.12$\pm$0.58 &  1.04$\pm$0.32  &      171$\pm$38   &507$\pm$22   &  14  \\
                    &           &  3.001  &   1.22$\pm$0.62  &  0.92$\pm$0.65 &  1.10$\pm$0.46  &      170$\pm$38    &503$\pm$22  &       \\
J1429+5447 & 6.119  &    -       &    -                       &    -                      &    -                        &    -   &-            & 12  \\
J1545+6028 & 5.794  &  4.152  &   2.32$\pm$0.35  &  2.03$\pm$0.23 &  0.65$\pm$0.21  &      191$\pm$36   &475$\pm$18    &  20  \\
                    &           &  3.616  &   2.03$\pm$0.49  &  1.05$\pm$0.50 &  0.79$\pm$0.42  &      160$\pm$35   &564$\pm$16   &        \\
J1602+4228 & 6.083  &    -       &    -                       &    -                      &    -                        &    -    &-           &  13  \\
J1609+3041 & 6.146  &  3.896  &   1.33$\pm$0.23  &  1.54$\pm$0.27 &  1.66$\pm$0.23  &      132$\pm$35  &391$\pm$16    &  12  \\
J1621+5155 & 5.637  &    -       &    -                       &    -                      &    -                        &    -   &-           &  20  \\
J1623+3112 & 6.254  &    -       &    -                       &    -                      &    -                        &    -    &-          &  12  \\
J2310+1855 & 5.956  &  4.244  &   1.86$\pm$0.35  &  0.98$\pm$0.21 &  1.90$\pm$0.25  &   221$\pm$34  &555$\pm$14      &  18  \\  
                      &         &   4.013  &  1.19$\pm$0.21   &    1.20$\pm$0.26   &   0.83$\pm$0.33  &  $<$165 & $<$388 & \\
\hline    
\end{tabular}\label{table_ew} 
\end{center}
\footnotesize{(1)Quasars. (2) Emission redshift of the quasars. (3) Absorption redshift of Mg~{\sc ii} systems, measurement errors of $z_{abs}$ are smaller than 0.001. (4) Equivalent widths of Mg~{\sc ii} ($\lambda$2796) lines, which are from a Voight profile. The errors are measured by method introduced in Section \ref{sec_alg}. Same for column (5) and (6). (5) Equivalent width of Mg~{\sc ii} ($\lambda$2803) lines. (6) Equivalent width of Fe~{\sc ii} ($\lambda$2600) lines. (7) Velocity width of Mg~{\sc ii} ($\lambda$2796) lines. Instrument broadening was removed. The error of $\Delta v$ was computed from the quadratic sum root of 1 $\sigma$ error of FWHM$_{\rm arc}$ and FWHM$_{\rm obs}$, which are FWHMs of arc files and observed absorption profiles.  (8) Observed velocity width of Mg~{\sc ii} ($\lambda$2796) lines.  (9) Mean S/N of the spectra. }

\end{table*}

\begin{figure}
%\plottwo{figures/mock_spectra.png}{figures/J1148_broadened_curve.png}
\gridline{\fig{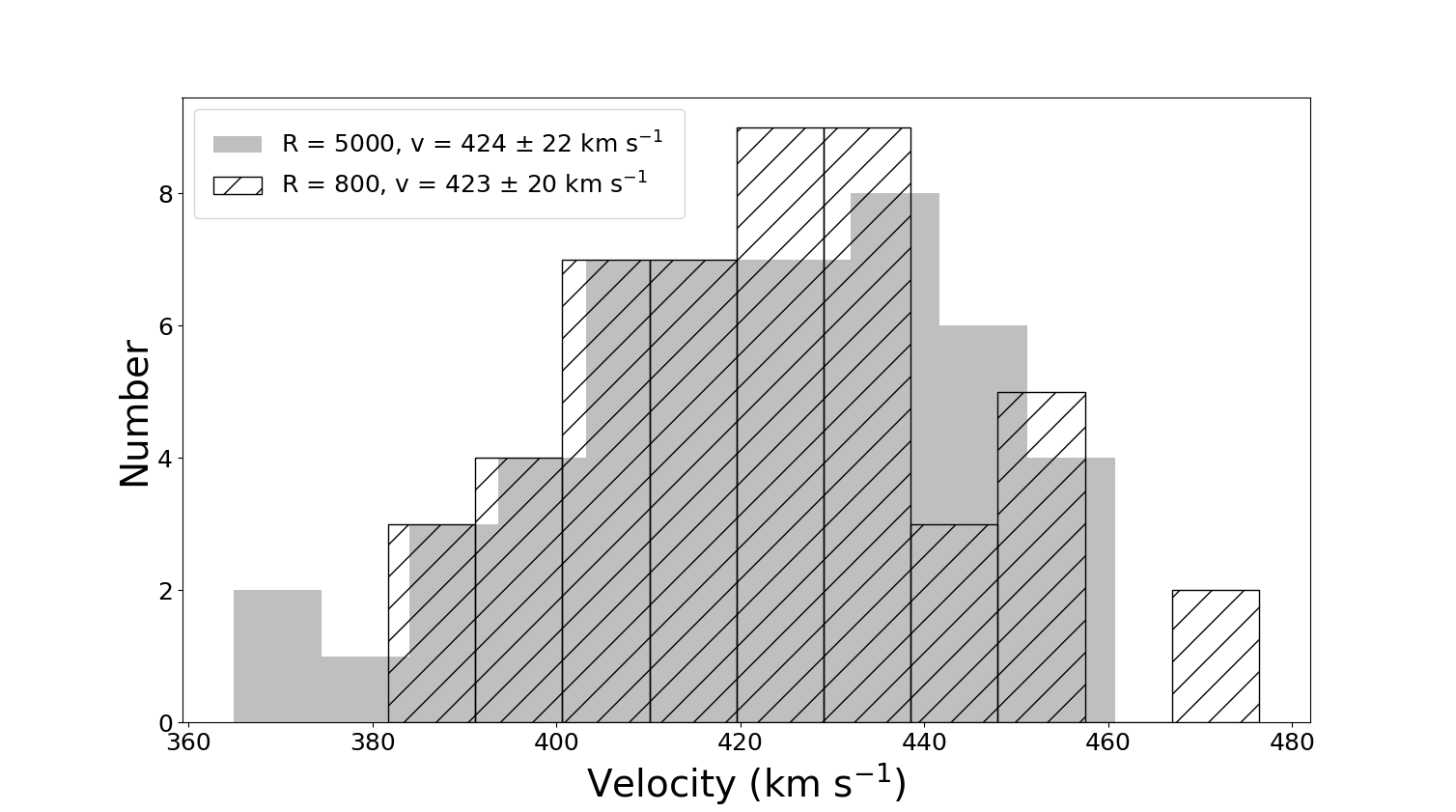}{0.5\textwidth}{(a)}}
\gridline{\fig{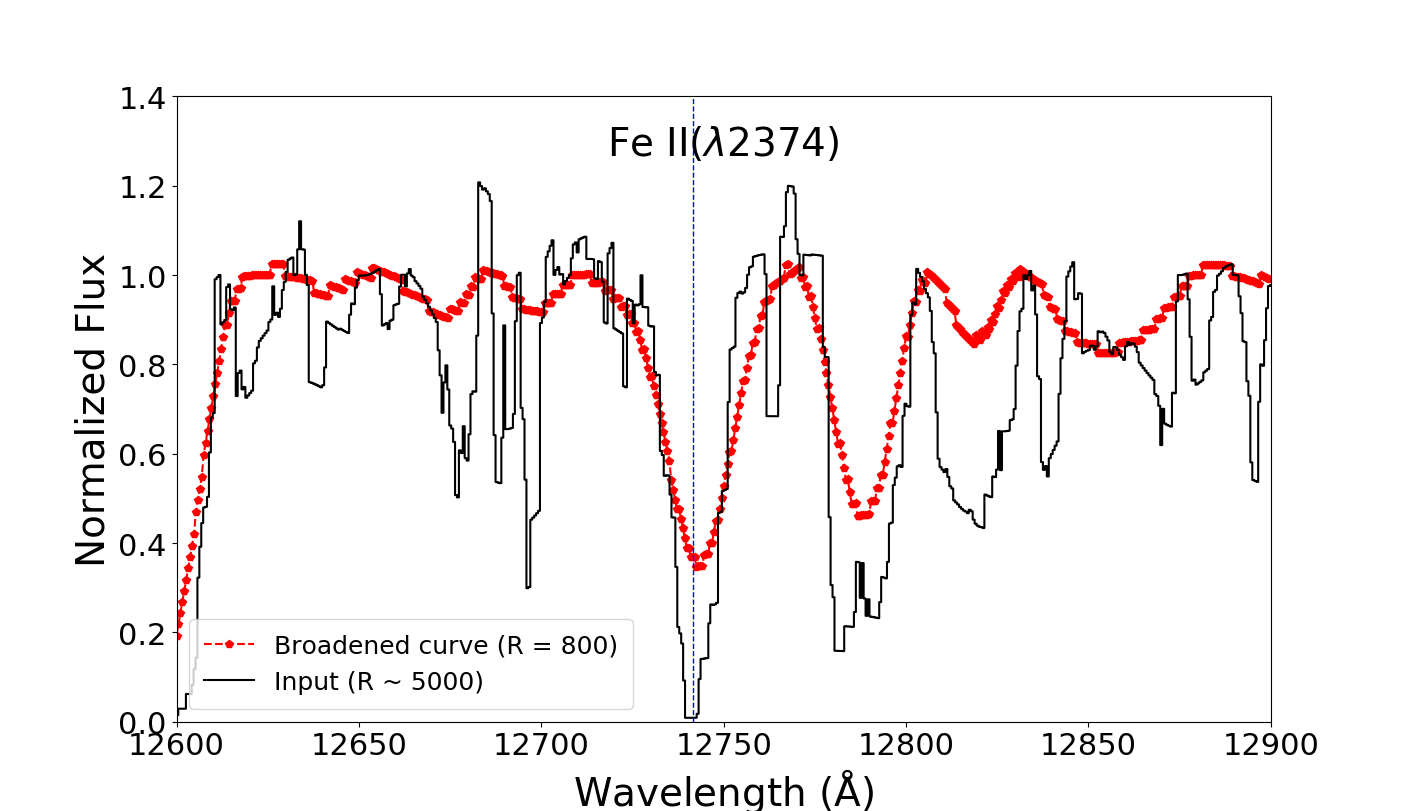}{0.5\textwidth}{(b)}}
%\gridline{\fig{figures/fire_gnirs_v_compare.png}{0.45\textwidth}{(c)}}
\caption{ (a). Velocity widths ($\Delta v$) histogram of 50 mock Mg~{\sc ii} doublets spectra with R = 800
(hashed) and R = 6000 (grey), respectively. Definition and measurement of $\Delta v$ are described in Section \ref{sec_mes}. The R = 6000 spectra are Voigt profiles with FWHM = 270 km s$^{-1}$. Then they were convolved into R = 800. Noise was added with a normal distribution and S/N was set to 10. The median $\Delta v$ measured with the method described in Section \ref{sec_mes} for R = 800 and R = 6000 spectra are 423 $\pm$ 20 km s$^{-1}$ and 424 $\pm$ 22 km s$^{-1}$. (b). Fe~{\sc ii} ($\lambda$2374) line at $z$ = 3.495 towards J1148+0702 from a Megellan--FIRE Spectrum. The input spectral resolution is around 6000 (black curve). The broadened red curve is the resolution--convoluted spectra with R$ = 800$. Velocity widths measured with input and broadened spectra are 566 km s$^{-1}$ and 531 km s$^{-1}$, respectively. 
}\label{fig:mock_spectra}
\end{figure}

\begin{figure}
  \resizebox{\hsize}{!}{\includegraphics{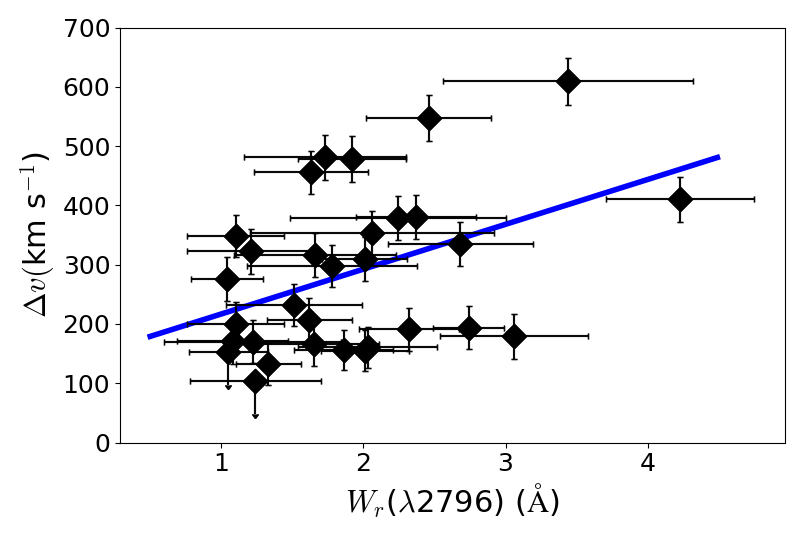}}
  \caption{\small{ Mg~{\sc ii} ($\lambda$2796) velocity width ($\Delta v$) against rest-frame equivalent width ($W_r$). The blue line is the linear relation of our sample with 2$\sigma$ limit: 75.63 km s$^{-1}$ \AA$^{-1} \times W_r$ + 141.19 km~s$^{-1}$ in equation \ref{eq:w_v}. 
}}\label{fig:v_w}
\end{figure}

%\begin{figure}
%  \resizebox{\hsize}{!}{\includegraphics{figures/fire_gnirs_v_compare.png}}
%  \caption{\small{ Velocity widths comparison of the overlapping sightlines (J0203+0012, J0836+0054,
%J0842+1218, J1148+0702 and J2310+1855) between this work and C17. Measurements in C17 with \wmgone $<$  1 \AA\ are not included. The gray deashed line is y = x. 
%}}\label{fig:v_comparison}
%\end{figure}

\subsection{Measurements}\label{sec_mes}
We also measured $W_r$ of an absorption candidate from a Voigt profile fit. The line is fitted using the VoigtFit package \citep{kro18}. During the visual inspection process, we noticed that our detection algorithm detected a few absorbers as candidates but they are strongly blended,  e.g. Mg~{\sc ii} ($\lambda$2803) lines at $z$ = 3.059 (J0002+2550), $z$ = 5.595 (J0840+5624), $z$ = 4.201 (J1250+3130) and $z$ = 4.530 (J1335+3533). Due to this blending, the $W_r$ would be overestimated from the flux boxcar summation. The Doppler parameter $b$ values of the fits are between 20--60 km s$^{-1}$.  Because that the relatively low resolution would introduce large uncertainties on the $b$ and column density measurements, we only use Voigt fits to calculate the $W_r$, which is independent of spectral resolution. We compared the measurements from the fits and the flux summation. Except for systems with obvious blending, the differences between the two measurements have a median of 0.13 \AA\ and a maximum of 0.5 \AA.

We then measured the velocity width from the best-fitted parameters. The intrinsic rest-frame velocity width $\Delta v$ was determined by the instrument broadening and observed velocity width $\Delta v_\mathrm{obs}$. The observed $\Delta v_\mathrm{obs}$ was measured between the leftmost and rightmost pixels with optical depth $\tau <0.1$. The optical depth $\tau$ equals ln (1/$F$), where $F$ is the normalized flux at this wavelength. This measurement is similar to the standard $\Delta v_{90}$ definition \citep{pro06}. The idea is to include all satellite absorption and to have a good representation of the kinematic extent of the absorption. We measured instrument broadening from lamp/arc lines used for wavelength calibration. The average FWHM of the arc lines FWHM$_{\rm arc}$ is roughly 376 $\pm$ 31 \kms (with 1 $\sigma$ error). The rest-frame intrinsic FWHM was calculated by ${\rm FWHM} = \sqrt{{\rm FWHM_{obs}}^2 - {\rm FWHM_{arc}}^2}$ / ($1+z$), where FWHM$_{\rm obs}$ is the observed FWHM of the line. Then we assume the ratio of intrinsic FWHM and $\Delta v$ is the same as the ratio of observed FWHM and $\Delta v_\mathrm{obs}$, i.e. $\Delta v$ = FWHM$\times(\Delta v_\mathrm{obs}/ {\rm FWHM_{obs}})$. 

To minimize the low resolution impact on our velocity spread measurements, we create 50 mock Mg~{\sc ii} absorption spectra and convolved them into FIRE resolution of R = 6000 (i.e. 50 km s$^{-1}$) and GNIRS resolution of R $\sim$ 800 (i.e. 376 km s$^{-1}$), respectively. Then we measured the $\Delta v$ from the mock spectra with different resolutions using the same method described above. The measurements are consistent within errors $\sim$ 20 km s$^{-1}$ (see Figure \ref{fig:mock_spectra}). We also used a FIRE spectra of Fe~{\sc ii} ($\lambda$ 2374) system at $z = $ 3.495 towards QSO J0148+0702. We degraded the spectral resolution into R = 800. The velocity width measurement difference between the original and degraded spectra is within 30 km s$^{-1}$. 

The intrinsic and observed velocity widths of all the detected absorbers are shown in Table \ref{table_ew} column (7) and (8), respectively. We found that 15 out of 32 absorbers have $\Delta v > 300$ km s$^{-1}$.  We fit the relation between $\Delta v$ and $W_r$ using a polynomial curve fitting technique considering the errors from two variables (see Figure \ref{fig:v_w}). 
\begin{equation}\label{eq:w_v}
\Delta v =  75.63 \: \mathrm{km~s^{-1}}  {\textrm \AA^{-1}} \times W_r + 141.19\: \mathrm{km~s^{-1}}.
 \end{equation}

\subsection{Comparison with C17}
We compared measurements of five overlapping sightlines (J0203+0012, J0836+0054, J0842+1218, J1148+0702 and J2310+1855) between our sample and the FIRE sample in C17. The details are presented in the following and Table \ref{table:c17compare}.

All \Mgtwo systems in J0203+0012 and the one at $z$ = 2.299 toward J0836+0054 reported in C17 are below 1 \AA, which are beyond our detection limit. The $W_r$, $\Delta v$ and $z$ measurements of the system at $z$ = 3.745 toward J0836+0054 are consistent. For J0841+1218, three systems were detected in this work and C17 at $z = 5.050, 2.540, 2.392$. The $W_r$, $\Delta v$ and $z$ measurements are consistent with errors. For J1148+0702, we detected two systems at $z$ = 4.369 (\wmgone = 4.23 $\pm$ 0.52 \AA) and $z$ = 3.495 (\wmgone = 6.50 $\pm$ 1.20 \AA). The measurements of the system at $z$ = 4.369 are consistent with that in C17.  The system at $z$  = 3.495 has extremely large velocity width ($>$ 800 km s$^{-1}$ for Mg~{\sc ii} ($\lambda$2796) line) and the absorptions are strongly blended within the doublet. Thus, the $W_r$ and $\Delta v$ measurements of this system inevitably have large uncertainties. We did not use this measurement when calculating the relation in Equation \ref{eq:w_v}. For J2310+1855, the two systems at $z$ = 3.299 and 2.351 detected in C17 have $W_r <$ 1.0 \AA, which are beyond our detection ability. The one at $z$ = 2.243 is located in a noisy region where the line are not able to be detected in our spectrum. We detected two systems at $z$ = 4.244 and $z$ = 4.013, which are not included in C17. The first one has \wmgone = 1.86 $\pm$ 0.35 \AA\ and $\Delta v$ = 221 $\pm$ 34 km s$^{-1}$. The second one has \wmgone = 1.19 $\pm$ 0.21 \AA\ and $\Delta v <$ 165 km s$^{-1}$ (see the last two panels in Figure \ref{fig:wmg_vmg}). The system at $z$ = 4.244 was present in an inspection of the spectrum used in C17, but was rejected by their automated search algorithm because \wmgtwo $>$ \wmgone, likely because of blending in the Mg~{\sc ii} ($\lambda$2803) line from interloping systems at lower redshift. The system at $z$ = 4.013  has severe telluric noise in FIRE spectrum (R. Simcoe, private communication).

In summary, except for the systems have \wmgone $<$ 1 \AA\ or where the spectra S/N is too low, our absorption redshfit, equivalent width and velocity spread measurements are consistent with that in C17 within errors. Though it is possible that, for systems have $\Delta v_\textrm{obs} <$ 400 km s$^{-1}$ (close to GNIRS resolution), our velocity widths uncertainties would be large.
%J0836+0054Our measurement are \wmgone = 2.46 $\pm$ 0.44 \AA\ and $\Delta v$ = 548 $\pm$ 39 km s$^{-1}$. The measurement in C17 of this system are \wmgone = 2.509 $\pm$ 0.016 \AA\ and $\Delta v$ = 510.4 km s$^{-1}$.

%J0842+1218 Our measurements for the \Mgtwo system at $z$ = 5.050 are \wmgone = 1.66 $\pm$ 0.57 \AA\ and $\Delta v$ =  301 $\pm$ 37 km s$^{-1}$. Measurements in C17 are \wmgone = and 1.81 $\pm$ 0.15 \AA\ and 245.1 km s$^{-1}$. Our measurements for the \Mgtwo system at $z$ = 2.540 are \wmgone = 2.68 $\pm$ 0.51 \AA\ and $\Delta v$ =  310 $\pm$ 37 km s$^{-1}$. Measurements in C17 are \wmgone = and 2.16 $\pm$ 0.10 \AA\ and 384.5 km s$^{-1}$. Our measurements for the \Mgtwo system at $z$ = 2.392 are \wmgone = 2.01 $\pm$ 0.30 \AA\ and $\Delta v$ =  279 $\pm$ 37 km s$^{-1}$. Measurements in C17 are \wmgone = 1.44 $\pm$ 0.25 \AA\ and 193.8 km s$^{-1}$. 

%The $\Delta v$ we measured for these two systems are 410 and 865 km s$^{-1}$. Equivalents widths measurements reported in C17 are \wmgone = 4.78 $\pm$ 0.11 \AA\ and \wmgone = 4.82 $\pm$ 0.19 \AA\, respectively. Velocity widths measurements in C17 are 371.9 km s$^{-1}$ and 899.2 km s$^{-1}$. 

%---------------------------------------------------------------------------------------------------------------------------------------
\section{Results}\label{sec_stat}

\begin{table*}
\begin{center}
  \caption{Measurements of the overlapping sightlines between this work and C17. J0203+0012 is not in this table, because the $W_r$ measurements of all systems detected in C17 are beyond our detection limit.}
\begin{tabular}{cccccc}
\hline
  (1) Quasar  & (2) $z_{abs}$              & (3) $W_r$($\lambda$2796) GNIRS   & (4)$W_r$($\lambda$2796) C17 & (5)$\Delta v$ GNIRS & (6) $\Delta v$ C17 \\
  &&(\AA)&(\AA)&(km s$^{-1}$)&(km s$^{-1}$) \\
\hline
    J0836+0054 &3.745 &2.46$\pm$0.44 &2.51$\pm$0.02& 548$\pm$39 & 510.4 \\
    J0842+1218 &5.050 &1.66$\pm$0.57 &1.81$\pm$0.15&301$\pm$37    &245.1\\
                         &2.540 &2.68$\pm$0.51 &2.16$\pm$0.10 & 310$\pm$37  &384.5\\
                         &2.392 &2.01$\pm$0.30 &1.44$\pm$0.25 &279$\pm$37   &193.8 \\
    J1148+0702 &4.369 & 4.23$\pm$0.52 & 4.78$\pm$0.11 & 410$\pm$38 &371.9\\
                          & 3.495&6.50$\pm$1.20 &4.82$\pm$0.19 & $>$865 &899.2\\   
   J2310+1855 & 4.244  & 1.86$\pm$0.35 &&      221$\pm$34   & \\
                            &4.013 & 1.19$\pm$0.21 &&  $<$ 165     & \\
\hline
\end{tabular}\label{table:c17compare}
\end{center}
\footnotesize{(1) Quasars. (2) Mg~{\sc ii} absorption redshifts. (3) Rest-frame equivalents of Mg~{\sc ii} ($\lambda$2796) measured in GNIRS. (4) Rest-frame equivalents of Mg~{\sc ii} ($\lambda$2796) in C17. (5) Velocity widths measured in GNIRS. (6) Velocity widths measured in C17. }  
\end{table*}

The near-IR spectra are strongly contaminated by OH skylines and telluric absorption. To conduct population statistics analysis for absorbers, we need to correct the incompleteness caused by the contamination. In this section, we first correct these effects and then present the statistical $dN/dz$ and $dN/dX$ for our \Mgtwo sample.

\subsection{Completeness}

\begin{figure}
  \resizebox{\hsize}{!}{\includegraphics{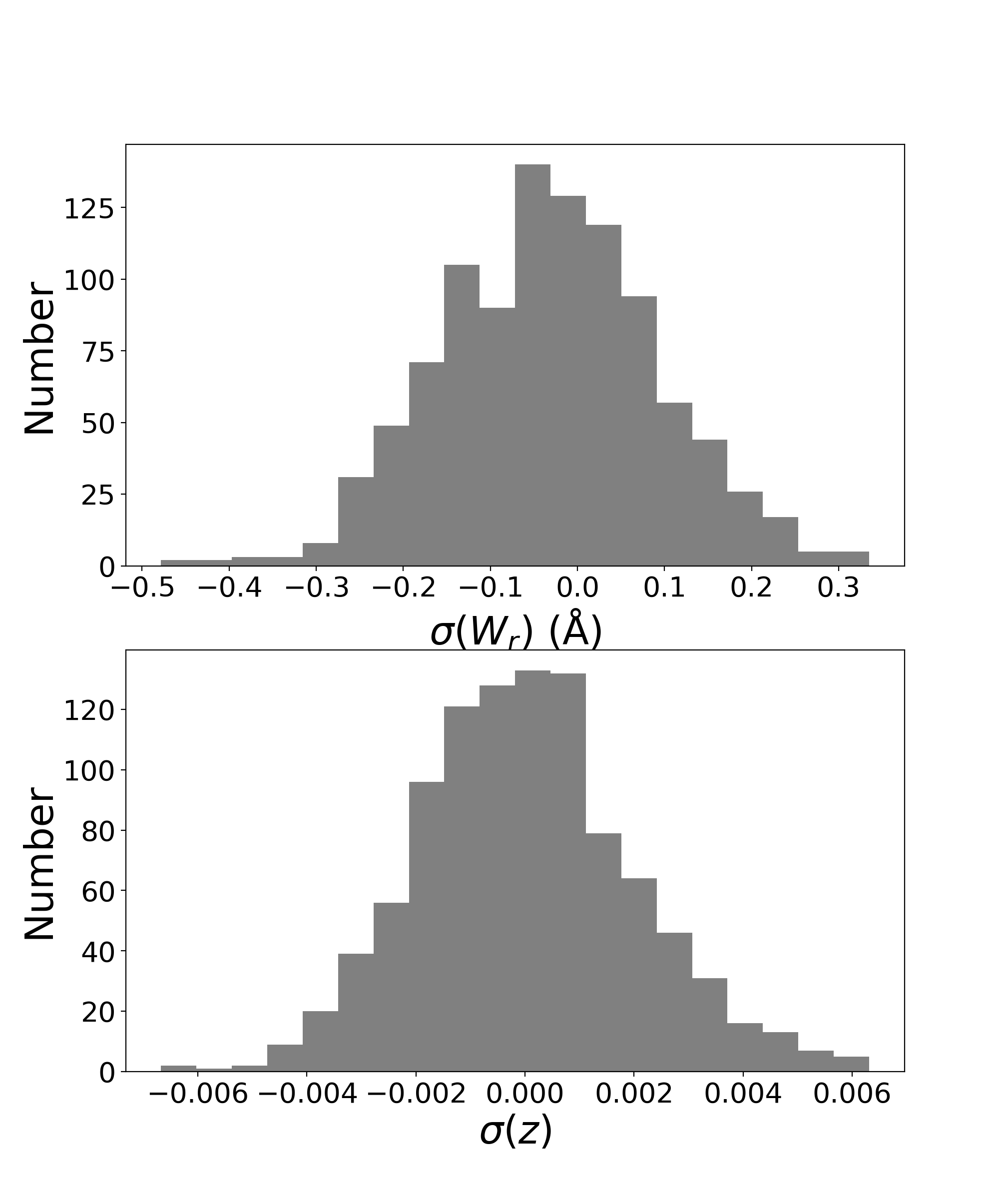}}
  \caption{\small{Measurement errors of \wmgone and redshift ($z$) from our simulation. The errors of $W_r$ and $z$ are 0.15 \AA and 0.002, respectively. The errors were calculated by comparing the inserted and measured values of 1000 mock Mg~{\sc ii} absorbers.   
}}\label{fig:w_z_errors}
\end{figure}

\begin{figure}
  \resizebox{\hsize}{!}{\includegraphics{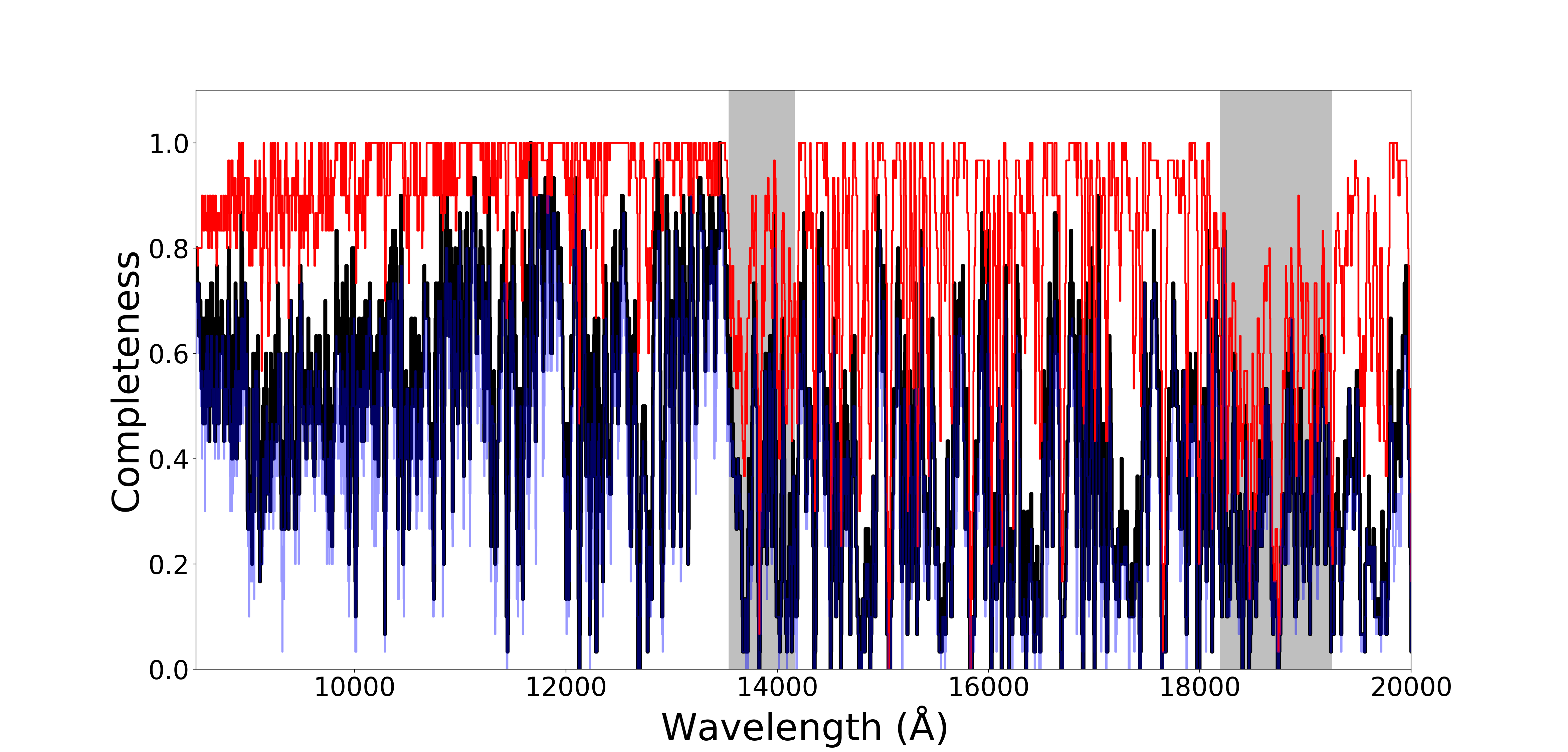}}
  \caption{\small{Average pathlength-weighted completeness $\overline{C}(W_r,z)$ for 31 sightlines in the sample with $W \geq 0.8$ (light blue), 1 (black) and 3 \AA~ (red), respectively. Two strong water vapor regions are labeled with vertical grey regions. 
}}\label{fig:completeness}
\end{figure}

\begin{figure}
  \resizebox{\hsize}{!}{\includegraphics{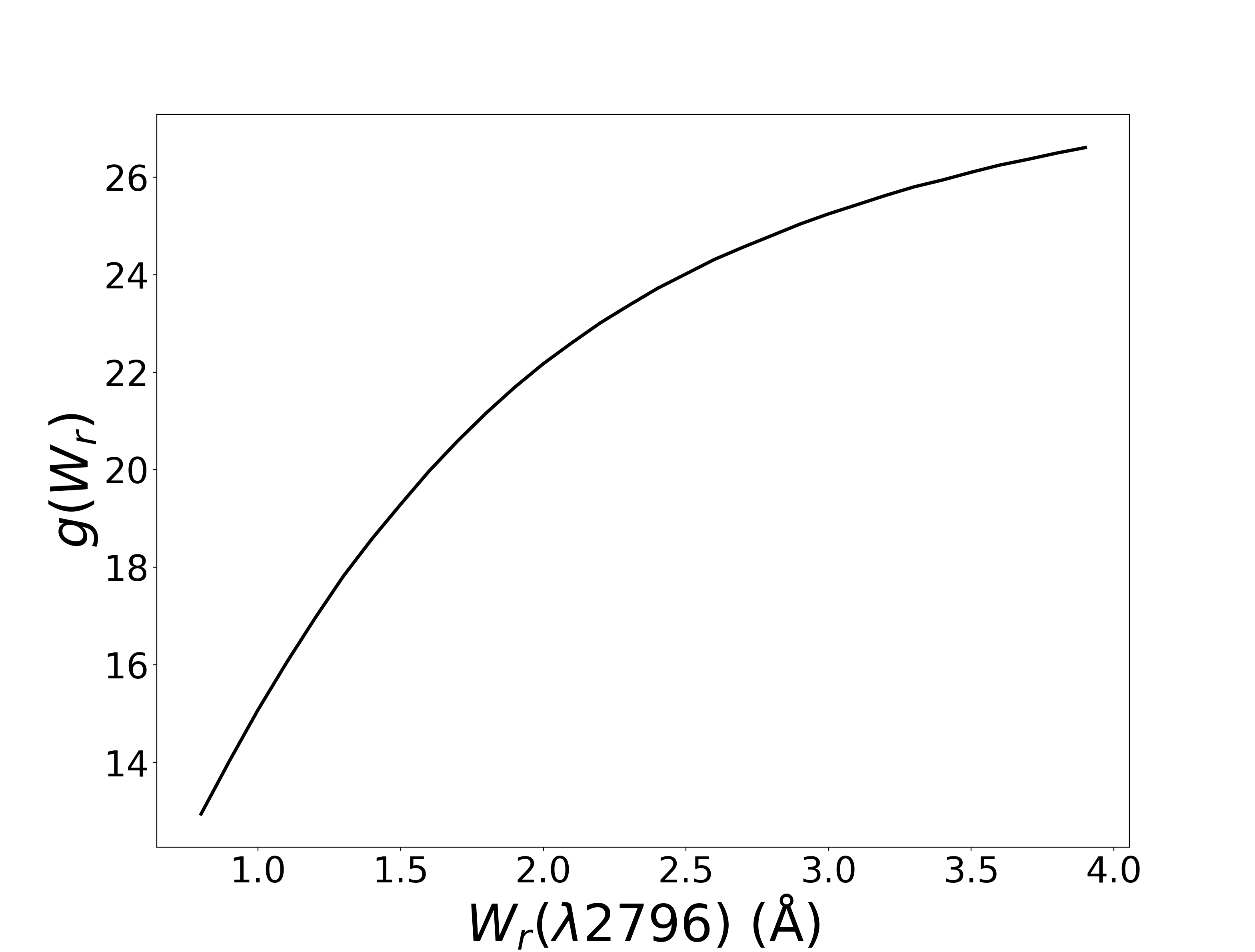}}
  \caption{\small{The total absorption path $g(W_r)$ against rest-frame equivalent width $W_r$. 
}}\label{fig:g_w}
\end{figure}

For each quasar spectrum, we performed a Monte Carlo simulation by inserting uniformly distributed, virtual \Mgtwo doublets in the wavelength range between 8500 \AA~ and  20,000\AA, corresponding to the absorber redshift  of 2.0 and  6.2, respectively. The inserted \wmgone varies between 1.0 and 4.5 \AA, which is the observed \w range of our detected \Mgtwo absorbers (except for the strongly blended one toward J1148+0702). For each $W_r$, its velocity width follows the relation $\Delta v$ = 103.37 km s$^{-1}$ \AA$^{-1}\times W_r$ + 399.60 km s$^{-1}$, which we measured from the $\Delta v_\textrm{obs}$ and $W_r$. Two strong water vapor regions (1.35 - 1.42 $\mu$m between $J$ and $H$ and 1.82 - 1.93 $\mu$m between $H$ and $K$ band) are discarded in the statistical analysis of $dN/dz$ and $dN/dX$. Then we use the algorithm introduced in Section \ref{sec_alg} to detect the inserted virtual absorbers. We measured the uncertainties between inserted and retrieved measurements from 1000 mock inserted Mg~{\sc ii} systems. The measurement errors of $W_r$ and $z$ are 0.015 \AA\ and 0.002, respectively (see Figure \ref{fig:w_z_errors}). This bias would be affect the final completeness significantly. The detection result is denoted as a Heaviside function $H(z, W_r)$:

\begin{equation}
	    H(z,W_r) = \left\{
	    \begin{array}{cl}
		1,& \text{if the absorber is detected},\\
		0,& \text{if the absorber is not detected}.
		\end{array}
		\right.
\end{equation}
The redshift-weighted density $g(z,W_r)$ is a function of $W_r$ and $z$ denoted as
\begin{equation}
g(z,W_r) = \sum^{N}_{i=1} H(z,W_r),
\end{equation}
where $N$ is the total number of sightlines. The total path $g(z)$ is obtained as the integral of the path density over the whole range that we selected (see Figure \ref{fig:g_w}): 
\begin{equation}
g(z)  =\int_{W_0}^{\infty} g(z,W_r)dz,
\end{equation}
where $W_0$ is the \w limit. For each sightline, its completeness is the detection rate of the inserted absorbers. The completeness of pathlength is a function of redshift and $W_r$, 
\begin{equation}
C(z,W_r) = g(z,W_r)/N.
\end{equation}
We show the pathlength-averaged completeness $\overline{C}(z,W_r)$ for the selected 31 sightlines with \wmgone $>$ 0.8 \AA, 1 \AA\ and $>$ 3 \AA\ in Figure \ref{fig:completeness}. For \wmgone$>$ 1 \AA\, the completeness is around 40\% $\sim$ 80\% in the $J$ band (1.17--1.37 $\mu$m), and around 20\% $\sim$ 60\% in the $H$ band (1.49--1.80 $\mu$m). The low completeness in the $H$ band is due to the contamination of strong sky lines.

Even if the S/N of a spectrum is high enough to detect weak lines, visual inspection may miss some weak absorbers. The probability for users to confirm true absorbers is defined as user acceptance.  In M13 and C17, the user acceptance rate is defined as a function of S/N and has been considered in their calculations. M13 suggests that when S/N $>$ 10, the user acceptance is close to 1 and the rejection fraction of false-positive candidates is close to 0. 

\citet{zhu13a} identified 40,000 Mg~{\sc ii} absorbers from SDSS at 0.4 $<z<$ 2.3. By requiring the simultaneous detection of Fe~{\sc ii} lines  ($\lambda\lambda$ 2344, 2383, 2586, 2600) for each Mg~{\sc ii} absorption line, they recovered close to 100\% of strong absorbers in the Pittsburgh catalogs \citep{qui11}. In this work, we focus on the strong system with \wmgone $\ge1$ \AA\, for which we also confirm detection of Fe~{\sc ii} candidate lines at the same absorption redshift. Given this and that our database consists of spectra with S/N $>$ 10, we assume that our visual inspection is correct at a rate of ca 95\%. In the case that one or two Fe~{\sc ii} candidate lines reside in the water vapor region and are affected significantly by OH lines, the bias would be within this rate given the wide wavelength coverage of Fe~{\sc ii} candidate lines. We calculated the Fe~{\sc ii} lines association with strong Mg~{\sc ii} systems in M13. We found that there are 35 out of 37 (94.5\%) strong Mg~{\sc ii} systems (\wmgone $>$ 1 \AA) associated with at least three clear Fe~{\sc ii} lines. Additionally, spurious detection caused e.g. by C~{\sc iv} doublets $W_r(\lambda$1548  $>$ 0.5 \AA) are not detected in our spectra. Therefore, the false positive detection rate from the weaker lines is close to 0.

%and our selected spectra have S/N $>$10, so we assume that our visual inspection is 100\% correct. 

%{\bf This is because the detections of strong Mg~{\sc ii} candidates are strengthened by Fe~{\sc ii} lines at same redshfit. This method is used in

\subsection{$dN/dz$ and $dN/dX$}

\begin{figure}
   \resizebox{\hsize}{!}{\includegraphics{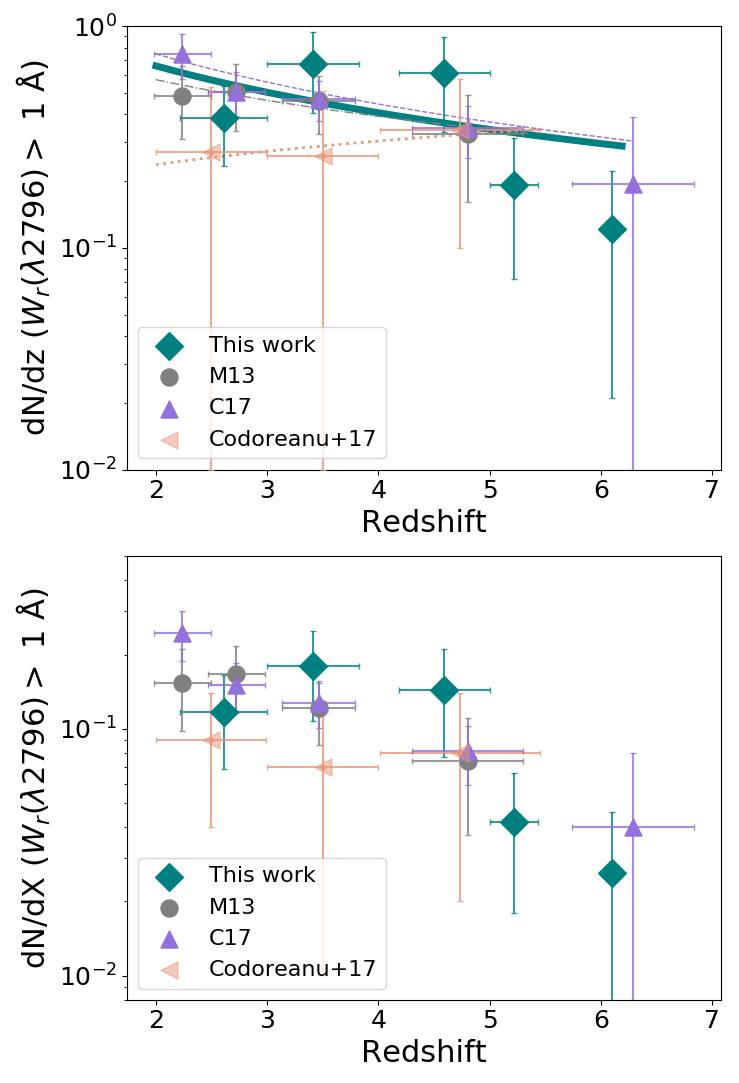}}
  \caption{\small{Line density $dN/dz$ (upper panel) and comoving line density $dN/dX$ (lower panel) of the \Mgtwo absorbers in our sample. Our results are plotted as green curves and diamonds. The purple triangles are the data from \citet{sfs17} and the grey dots represent data of \citet{mat12}. The orange triangles represent a sample of \citet{cod17}. The relation presented by M13 is (in purple) $dN/dz$ = (1.301$\pm$1.555)$\times$(1+$z)^{-0.746\pm0.857}$. Relation in C17 is (in grey) $dN/dz$ = (2.298$\pm$1.561) $\times$(1+$z)^{-1.020\pm0.475}$. The orange dashed line is relation in \citet{cod17}: $dN/dz$ = (0.14$\pm$0.09)$\times$(1+$z)^{0.48\pm0.20}$.
}}\label{fig:dndz_compare}
\end{figure}

We calculate the incompleteness-corrected line-of-sight density of strong \Mgtwo absorbers at different redshift bins. The results at four redshift bins between 2.2 and 6.0 are shown in Figure \ref{fig:dndz_compare} and Table \ref{table:dndz}. The relation between $dN/dz$ and redshift can be expressed as,
\begin{equation}
    \frac{dN}{dz} = N_0\times(1+z)^{\beta},
\end{equation}
where $N_0$ is the normalization and $\beta$ is the slope. We apply the Maximum Likelihood Estimation (MLE) method to the relation and find that $N_0$ and $\beta$ are 1.882$\pm$3.252 and --0.952$\pm$1.108, respectively. 

Previous studies (e.g., M13 and C17) have found that the $dN/dz$ of strong \Mgtwo absorbers generally decreases with increasing redshift at $2< z <6$. In particular, the $dN/dz$ or $dN/dX$  at $z >4.5$  drops rapidly (see Figure \ref{fig:dndz_compare}).  \citet{cod17} studied \Mgtwo systems using four quasars from VLT-Xshooter and found that the $dN/dz$ is relatively flat at $2<z\le4$. This is likely due to the larger uncertainties from their small sample, as they have pointed out in the paper. As shown in Figure \ref{fig:dndz_compare}, our results are consistent with the previous results within errors. The trend at $2<z<4$ is not clear due to the large errors, but the density decreases significantly at $z>4.5$.

\subsection{Kinematics and saturation}

\begin{figure*}
  \resizebox{\hsize}{!}{\includegraphics{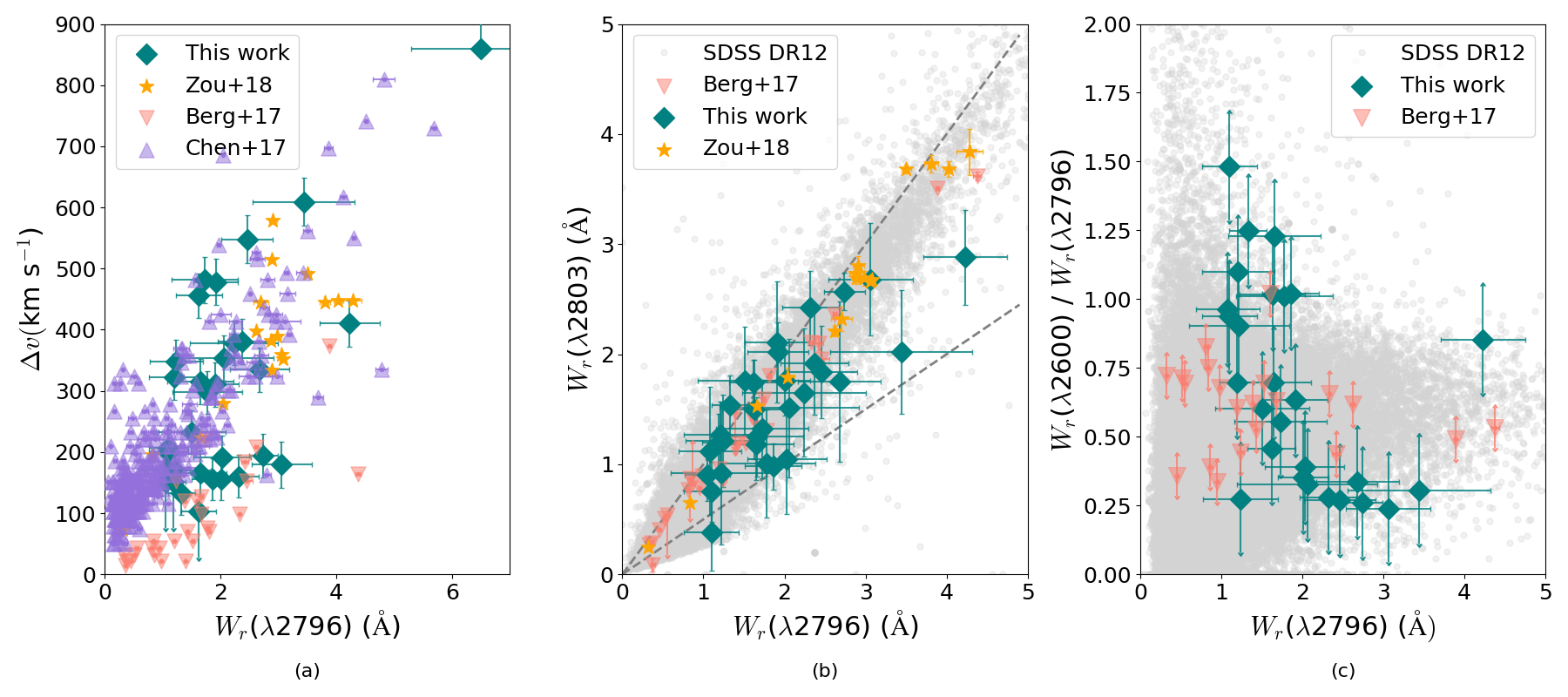}}
  \caption{Properties of the absorbers in our sample and comparison with previous studies: a SDSS sample of \Mgtwo systems $z<2$ from \citet{zhu13a}, a sample of DLA--\Mgtwo systems from XQ-100 survey \citep{berg17}, and a sample of Mg~{\sc ii} associated with neutral atomic carbon (C~{\sc i}) absorbers in \citet{zou18}. (a) Velocity width $\Delta v$ against \wmgone. The purple triangles are the data from C17. Our \Mgtwo systems are very likely affected by galactic superwinds. (b) \wmgtwo - \wmgone relation for different samples. The two dashed line represents the ratio of \wmgtwo/\wmgone = 1.0, 0.5, respectively. Our \Mgtwo lines exhibit potential less saturation than other two samples. (c) Ratio of \wfeone/\wmgone against \wmgone. Our strong \Mgtwo systems show relatively smaller \wfeone/\wmgone ratios, one possible reason is that \Mgtwo clouds at higher redshift  are in the interactions of multiple unresolved subcomponents.}\label{fig:mg_fe_compare}
\end{figure*}

The evolution of Mg~{\sc ii} incidence has implications for the origin of \Mgtwo absorbers. One possible scenario is that superwinds give rise to strong \Mgtwo absorbers in starburst galaxies \citep{bond01,heck01,bou06}. Superwinds are gas bubbles generated by starbursts. They escape from gravitational wells and then blow into galaxy halos.  Low-ions such as \Mgtwo and Na~{\sc i} reside in the shells of these superwinds. Another possible scenario is that strong Mg~{\sc ii} systems would reside in a galaxy groups environment. For example, \citet{gau13} find their ultra-strong Mg~{\sc ii} absorber (\wmgone = 4.2 \AA) at $z$ = 0.5624 is associated with five galaxies within 60 kpc.

%three, normal galaxies with relatively low SFR are found to have redshifts consistent with the absorbing gas.

\begin{table}
\begin{center}
  \caption{Pathlength density of \Mgtwo absorbers. }
\begin{tabular}{cccc}
\hline
  (1) $\Delta z$ & (2) $\overline{C} (\%)$              & (3) $dN/dz$  & (4) $dN/dX$  \\
\hline
             2.220 -- 3.000 & 47.1   &0.386$\pm$0.153 &   0.117$\pm$0.048   \\ 
             3.000 -- 3.828 & 49.6   &0.672$\pm$0.268 &   0.179$\pm$0.071   \\           
             4.185 -- 5.000 & 29.7   &0.612$\pm$0.280 &   0.144$\pm$0.067   \\ 
             5.000 -- 5.436 & 31.6   &0.192$\pm$0.120 &   0.042$\pm$0.024   \\ 
             6.000 -- 6.200 & 25.1   &0.121$\pm$0.100 &   0.026$\pm$0.020   \\ 
\hline
\end{tabular}\label{table:dndz}
\end{center}
\footnotesize{(1) Each redshfit bin selected to calculate the pathlength density. (2) Average pathlength-weighted completeness over 31 sightlines. (3) Number of absorbers per unit redshift $dz$. (4) Number of absorbers per comoving absorption distance $dX$.}  
\end{table}

To further investigate the possible scenarios for the origin of our strong \Mgtwo systems,  we compare the velocity widths of our \Mgtwo systems with those at similar and lower redshift. We compare with three samples in the literature: a blindly-searched Mg~{\sc ii} sample from SDSS DR12 \citep{zhu13a} at 0.4 $<z<$ 2.3, Mg~{\sc ii} systems associated with a DLA sample from the XQ-100 survey at 2 $<z<$ 4 \citep{berg17}, and Mg~{\sc ii} systems traced by a neutral atomic carbon (C~{\sc i}) sample at 1.5 $<z<$ 2.7 \citep{zou18}. The comparison is plotted in Figure \ref{fig:mg_fe_compare}. 

We found that the velocity widths of our \Mgtwo absorbers are larger than those associated with DLAs with similar equivalent widths at 2$<z<4$, this feature is also seen in C17 strong Mg~{\sc ii} systems. In the C17 sample of 287 absorbers, 104 of which have \wmgone $>$ 1 \AA, and 58 out of 104 have $\Delta v >$ 300 km s$^{-1}$. Note that the $\Delta v$ given in C17 is defined as the total velocity interval under the continuum. Even if only 90\% of their intervals are considered, half of the strong absorbers still have $\Delta v >$ 300 km s$^{-1}$. In the DLA-tracing Mg~{\sc ii} sample, 18 out of 29 Mg~{\sc ii} absorbers have $W_r >$ 1 \AA\, but only one has velocity width greater than 300 km s$^{-1}$. Moreover, large velocity widths for \Mgtwo absorbers are also seen in the C~{\sc i}-tracing Mg~{\sc ii} absorbers at 1.5 $<z<$ 2.7. The velocity widths were measured by the same method decribed in Setion \ref{sec_mes}. In the 17 systems of C~{\sc i}-tracing Mg~{\sc ii} absorbers, 15 of which have $W_r >$ 1 \AA\, and 13 out of 15 (87\%) have $\Delta v >$ 300 km s$^{-1}$. C~{\sc i} has been shown to effectively trace molecular and cold gas at $z \sim 2$, and thus star formation activities. As discussed in \citet{zou18}, the C~{\sc i} systems can be highly disturbed by superwinds or the interactions between several galaxies. Therefore, large velocity widths of our Mg~{\sc ii} absorber suggest that our systems are potentially strongly affected by the galactic superwinds and/or the interaction within galaxy groups.  

The two dashed lines in panel (b) are for \wmgtwo/\wmgone = 1 and 0.5 respectively. The ratio greater than one implies that \Mgtwo doublets are strongly saturated. In our sample, about 42\% of the absorbers have this line ratio greater than 0.8. This fraction is $\sim55$\% in the 0.4 $<z<$ 2.3 SDSS sample. Our \Mgtwo systems are slightly less saturated than the absorbers at $z<$ 2.3. 

Another piece of possible supportive evidence is the equivalent width ratio of Fe~{\sc ii} and Mg~{\sc ii} lines (\wfeone/\wmgone).
We compare our sample with the DLA-tracing Mg~{\sc ii} sample at 2$<z<4$ and a sample from \citet{hid12} at low redshift. \cite{hid12} analyzed 87 \Mgtwo system with \wmgone $>$ 0.3 \AA\ at 0.2 $< z <$ 2.5. They found that strong systems (\wmgone $>$ 1 \AA) do not have small \wfeone/\wmgone ratios in their sample. 
In panel (c) of Figure \ref{fig:mg_fe_compare}, our sample covers a wide range of the \wfeone/\wmgone ratios. In particular, four systems among the strongest Mg~{\sc ii} absorbers have smaller ratios ($W_r$ ($\lambda$2600)/$W_r$ ($\lambda$2796) $<$ 0.5) than most of the other systems in the sample. The small \wfeone/\wmgone values can be due to many reasons, e.g. kinematics evolution, dust depletion and intrinsic [Mg/Fe] abundance in the gas phase. We here propose that the kinematic evolution of the profiles of the very strong absorbers is a possible reason. Which means, at high redshift, the number of unresolved sub-components associated with strong Mg~{\sc ii} absorbers may grow.

%Additionally, {\bf nine systems have the ratio smaller than 0.36 (with a median of 0.28). 
%If we simply the convert the $W_r$ ratio to the abundance ratio, this value (0.36) is a [Fe/Mg] plateau reported in the star formation history model at [Fe/H] $<-$1 (see e.g. \citealt{kobayashi20}). The [Fe/Mg] ratio increases rapidly due to a delayed contribution of Type Ia SN at [Fe/H] $>$ --1. In the star formation model in \citet{kobayashi20} Figure 1, [Fe/H] starts to be greater than -- 1 after 3.5 Gyr ($z\sim$ 1.9). Therefore, the strong \Mgtwo with weak \Fetwo at $z> 2$ could partially be due to the nucleosynthetic effects. } 

\begin{table*}[!ht]
\begin{center}
 \caption{Photometry of possible galaxies counterparts around our targets selected. The selection criteria is $\Delta v>$ 300 \kms or  $W_r(\lambda$2796) $>$1.5 \AA. }
  \setlength{\tabcolsep}{.75mm}{
\begin{tabular}{lccccccccccc}
\hline
Quasar        & $z_{abs}$ &  Targets NO. &R.A.              & Dec.             & $F105W$      & $M$   & $D$  & $g$  & $i$  & $r$   &  $z$\\
 &&&&& (mag)&(mag)&(kpc) &(mag) &(mag)&(mag)&(mag)\\
\hline
J0002+2550 &  3.059    &   1          & 00:02:39.24    & +25:50:36.7    & 24.74$\pm$0.08 & --19.15$\pm$0.08 & 20.2  & $>$25.78   &&&\\
J0050+3445 &  3.435    &   1          & 00:50:06.99    & +34:45:22.8    & 24.83$\pm$0.08 & --19.20$\pm$0.08 & 29.7   & 25.44$\pm$0.10  & 25.64$\pm$0.11&& \\
                       &             &   2          & 00:50:06.71    & +34:45:18.5    & 23.96$\pm$0.04 & --20.07$\pm$0.04 & 30.8  &25.43$\pm$0.10  & 24.89$\pm$0.13&& \\
J0842+1218 &  5.050     &  1          & 08:42:29.55    & +12:18:52.5    & 24.72$\pm$0.08 & --19.72$\pm$0.08  & 17.1   &  $>$25.64 &&&\\
                      &  2.540    &   1          &     &                                          & 24.72$\pm$0.08 & --18.94$\pm$0.08  & 17.1   &  &&&\\
                      &  2.392    &   1          &   &                                             & 24.72$\pm$0.08 & --18.87$\pm$0.08  & 21.9   &  &&&\\
J1207+0630 &  3.808     &  1          & 12:07:37.61    & +06:30:11.3    & 24.83$\pm$0.12 & --19.31$\pm$0.12  & 21.5  &  $>$25.62& &&\\    
                      & 3.808    &  2          & 12:07:37.55    & +06:30:13.9    & 25.10$\pm$0.16 & --19.04$\pm$0.12 & 30.1&  &&& \\ 
J1250+3130  & 4.201     &   1          & 12:50:51.93    & +31:30:23.6    & 25.49$\pm$0.16 & --18.76$\pm$0.16 & 11.4 &  $>$25.61&  &&  \\
                      &  3.860    &   1          &    &     & 25.49$\pm$0.16 & --18.66$\pm$0.16 & 11.8  & && &\\
                      &  2.292    &   1          &   &    & 25.49$\pm$0.16 & --18.04$\pm$0.16 & 13.7  &   &&&\\
%J1335+3533 &  4.53     &   1          & 13:35:50.64    & +35:33:13.49    & 25.93$\pm$0.29 & -18.39$\pm$0.29 & 20.92   &    \\
J1335+3533 &  4.530     &   1          & 13:35:50.68    & +35:33:11.5    & 23.69$\pm$0.04 & --20.63$\pm$0.04 & 30.7   &&&  &     \\ 
J2310+1855 &  4.244     &   1          & 23:10:38.73    & +18:55:17.2    & 24.26$\pm$0.05 & --20.00$\pm$0.05 & 22.6  && &25.55$\pm$0.14&24.34$\pm$0.25\\
                    &  4.244     &  2          & 23:10:38.95    & +18:55:22.6    & 25.10$\pm$0.11 & --19.15$\pm$0.11 & 21.4 & &&26.18$\pm$0.25&\\
            %       &    4.013    &   1          &                         &                     &24.26$\pm$0.05  & --19.94$\pm$0.05 & 22.9 &&&&    \\
           %         &   4.103    &   2          &                       &                       & 25.10$\pm$0.11  & --20.10$\pm$0.11 & 21.7  & &&&     \\ 
                   \hline
\end{tabular}\label{table:hst} }
\end{center}
\footnotesize{The 1 or 2 labels in the third column are galaxy candidates number in Figures 8 and 9. Two  detection limit in g band for J0002+2550,
J0842+1218, J1207+0630, and J1250+3130 are measured from public DECaLs imaging data. The g; r bands magnitudes for J0500+3445
and r; z bands magnitudes for J2310+1855 are measured from CFHT--MegaPrime.
}
%\ hspace{-1cm}
%\hspace*{-3cm}

\end{table*}

%subsection{O~{\sc i}}
%An extensive study of O~{\sc i} $\lambda1302$ at  $z>$ 5 was made by \citet{bec11}, who detected 9 O~{\sc i} absorption systems %with 0.031 \AA\ $<W_r<$ 0.391 \AA\ at 5.3 $<z<$ 6.4. The line-of-sight number density is 0.25$^{+0.21}_{-0.13}$, which is %similar to the number density of DLA/sub-DLA at 3 $< z <$ 5. The nearly constant line density of low-ions is different from that of C~{\sc iv} at $z>$ 4.5.  We did not detect strong O~{\sc i} absorbers at $z>$ 5.5 in our sample, since our detection limit for O~{\sc i} $\lambda$1302 at $z\sim$ 5.91 ($\sim$ 9000 \AA) is $W_r$ = 0.40 \AA. \citet{bec11} reported four systems towards quasar J1148+5251 at $z$ = 6.0115, 6.1312, 6.1988, 6.2575 and one system towards quasar J1623+3112 at $z$ = 5.8415. The two sightlines are included in our sample. However, the largest $W_r$ measured in these five systems is 0.391 \AA, which is below our detection limit.

%---------------------------------------------------------------------------------------------------------------------------------------
\section{Discussion}\label{sec_disussion}

Our strong \Mgtwo systems exhibit large rest-frame velocity widths and potential less saturation, the \Mgtwo gas is potentially strongly affected by galactic superwinds or the interaction within galaxy groups. Previous studies suggest that both star-forming and passive galaxies may host Mg~{\sc ii} absorbers. Star-forming galaxies tend to host stronger absobers. \citet{zib07} studied 2,800 Mg~{\sc ii} systems having \wmgone $>0.8$ \AA\ at 0.37 $<z<$ 1 and associated galaxies within 20-100 kpc. They tentatively conclude that \wmgone $<$ 1.1 \AA\ systems are associated with passive galaxies, while \wmgone $>$ 1 \AA\ systems tends to associated with star-forming galaxies. \citet{lan14} selected 2,000 galaxy-Mg~{\sc ii} absorber pairs at $z < $ 0.5 and found that, within 50 kpc, strong absorbers tend to be associated with star-forming galaxies. In this section, we will investigate the gas properties of the \Mgtwo clouds, including gas cross-section, absorbing halo size, and the galaxy impact parameter.

\begin{figure*}[htbp]
\centering
\begin{minipage}[t]{0.3\linewidth}
\centering
\includegraphics[width=4cm]{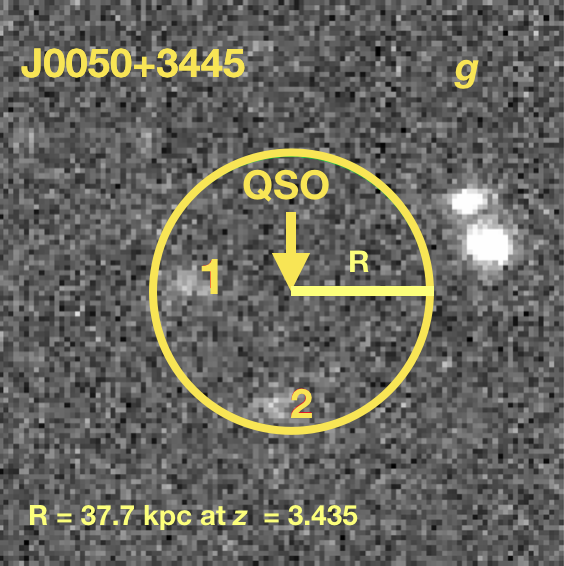}
%\caption{fig1}
\end{minipage}%
\begin{minipage}[t]{0.3\linewidth}
\centering
\includegraphics[width=4cm]{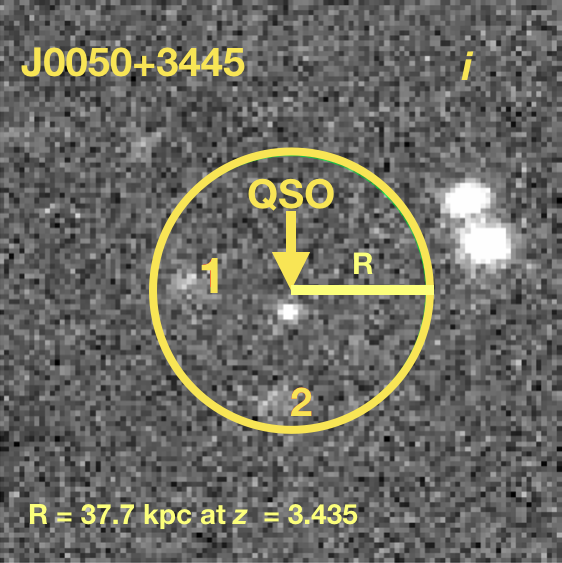}
%\caption{fig2}
\end{minipage}%
\begin{minipage}[t]{0.3\linewidth}
\centering
\includegraphics[width=4cm]{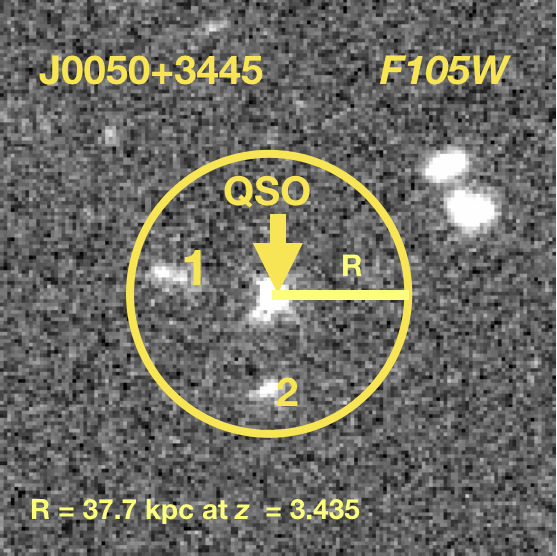}
%\caption{fig2}
\end{minipage}%
%%
%\caption{J0050+3445 images cutout in $g,i,F105W$ bands respectively.  Each figure size is 20$\arcsec\times$20$\arcsec$. The $g$ and $i$ band %images are from CFHT MegaPrime, the $F105W$ band image is from \it{HST}. The possible nearby galaxies are denoted as 1 and 2.}
%\label{fig:J0050_img}
%\end{figure*}

\begin{minipage}[t]{0.3\linewidth}
\centering
\includegraphics[width=4cm]{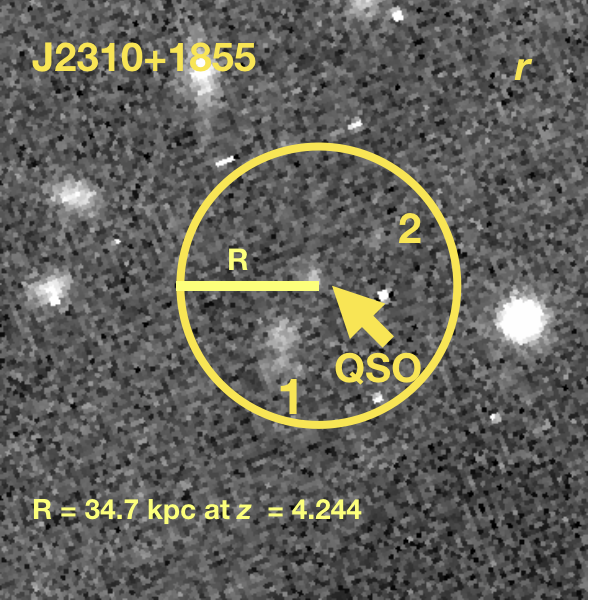}
%\caption{fig1}
\end{minipage}%
\begin{minipage}[t]{0.3\linewidth}
\centering
\includegraphics[width=4cm]{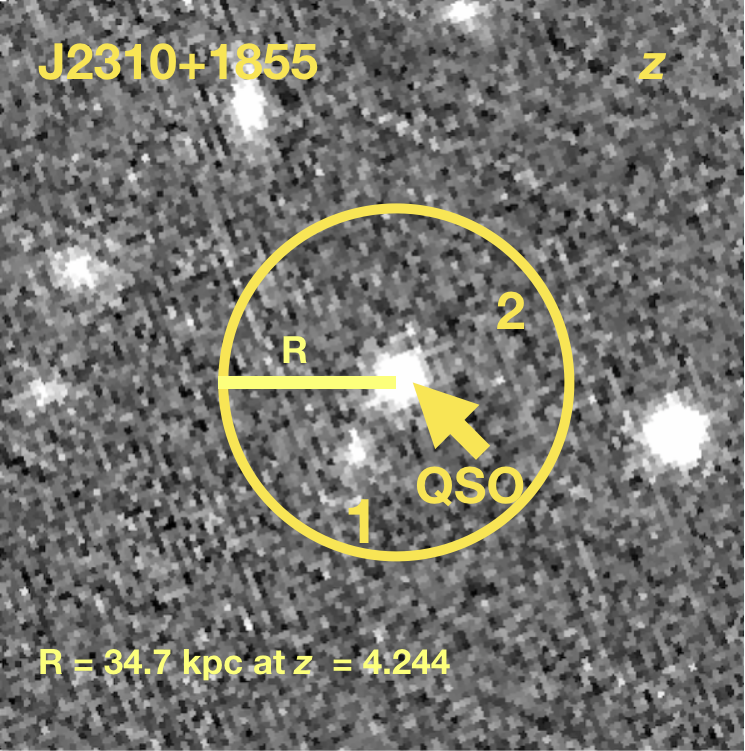}
%\caption{fig2}
\end{minipage}%
\begin{minipage}[t]{0.3\linewidth}
\centering
\includegraphics[width=4cm]{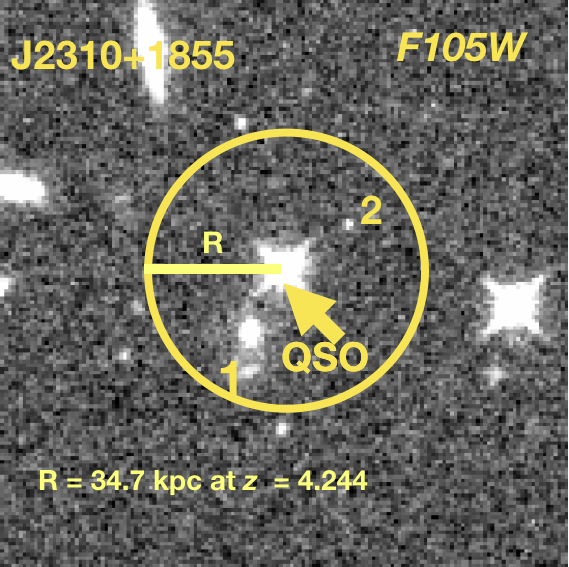}
%\caption{fig2}
\end{minipage}%
%
%\centering
\caption{Image cutouts of J0050+3445 and J2310+1855. The $g,i$ band images of J0500+3445 and $r,z$ band images of J2310+1855 are from CFHT MegaPrime. The $F105W$-band images are from {\it{HST}}. Image size is 20$\arcsec\times$20$\arcsec$, the yellow circle is in a 5 $\arcsec$ radius. The candidate absorber galaxies are denoted as 1 and 2.}
\label{fig:J0050_J2310_img}
\end{figure*}

\subsection{HST images}

\begin{figure*}[htbp]
\centering
\begin{minipage}[t]{0.3\linewidth}
\centering
\includegraphics[clip,width=4cm]{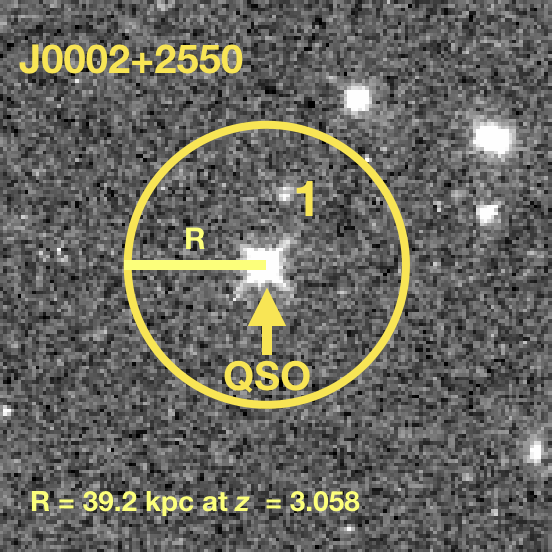}
%\caption{fig1}
\end{minipage}%
\begin{minipage}[t]{0.3\linewidth}
\centering
 \includegraphics[clip,width=4cm]{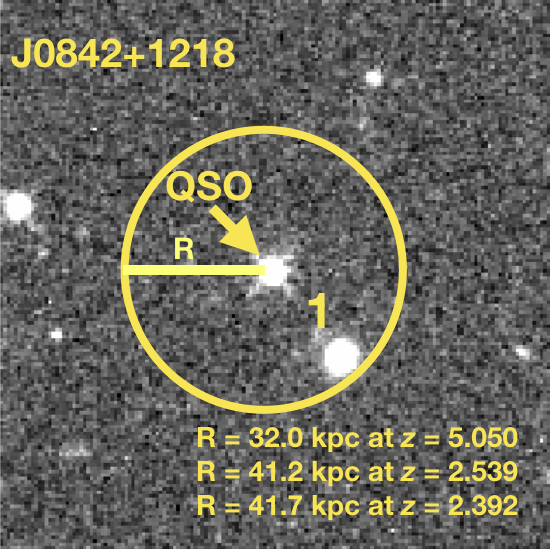}
%\caption{fig2}
\end{minipage}%
\begin{minipage}[t]{0.3\linewidth}
\centering
 \includegraphics[clip,width=4cm]{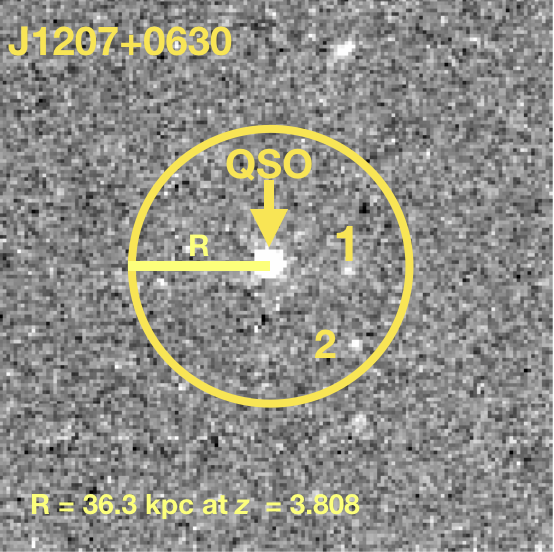}
%\caption{fig2}
\end{minipage}%

\begin{minipage}[t]{0.835\linewidth}
\flushleft
%\flushleft
\includegraphics[width=9.56cm]{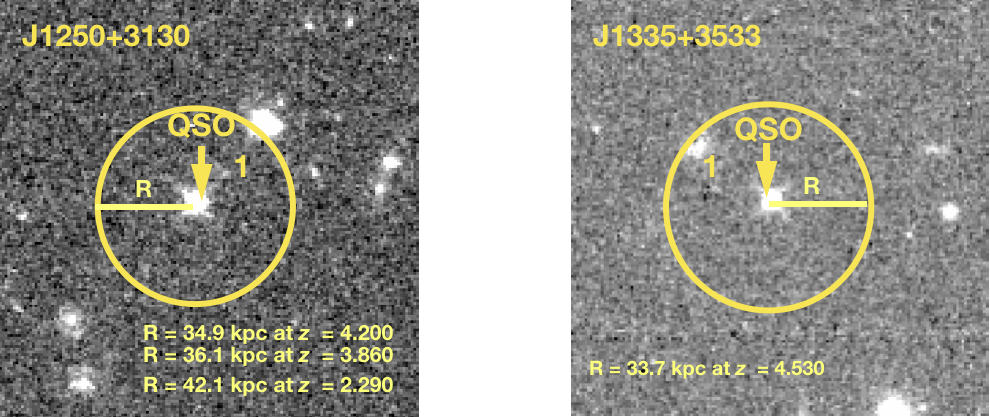}
%\caption{fig1}
\end{minipage}%
%
%\begin{minipage}[t]{0.45\linewidth}
%\flushleft%
%\includegraphics[width=4cm]{figures/J1335_hst.png}
%\caption{fig2}
%\end{minipage}%
%
%\begin{minipage}[t]{0.3\linewidth}
%\centering
%\includegraphics[width=4cm]{figures/J1429_hst.png}
%\caption{fig2}
%\end{minipage}%

%\begin{minipage}{0.828\linewidth}
%\flushleft 
 %\includegraphics[clip,width=4cm]{figures/J1602_hst.png}
%\end{minipage}
\caption{ {\it{HST}} $F105W$ band image cutouts for J0002+2550,  J0842+1218,  J1207+0630, J1250+3130,  J1335+3533,  J1429+5447 and J1602+4228. Each figure size is 20$\arcsec\times$20$\arcsec$, the yellow circle is in a 5 $\arcsec$ radius. The possible nearby galaxy are denoted as 1 or 2.} %There are two galaxy candidates within the 5 $\arcsec$ circle for J1429+5447 and one galaxy at 0.85$\arcsec$ offset from J1602+4228. However, no strong \Mgtwo or \Fetwo absorbers found for these two quasars.} %spectra. The figure size for each is 20$\arcsec \times 20 \arcsec$.}
\label{fig:hst}
\end{figure*}

We have a small sample of {\it HST} snapshot images observed in the WFC3/IR F105W band (Program ID: 12184, PI: X. Fan). The images cover 7 of our quasars with strong absorbers with \wmgone $>$ 1.5 \AA\ or $\Delta v >$ 300 km s $^{-1}$, namely J0002+2550, J0050+3445, J0842+1218, J1207+0630, J1335+3533, J2310+1855. The median redshift of the absorbers is 3.982. For each quasar, there is at least one candidate galaxy with $\ge$7$\sigma$ detection ($\le$ 25.5 mag) within a 5$\arcsec$ circular radius in the {\it HST} image (see Table \ref{table:hst} and Figures \ref{fig:J0050_J2310_img}, \ref{fig:hst}). The photometry is performed using SExtractor \citep{ber96}. All these galaxies are fainter than 24 mag in F105W, which is fainter than $L^*$ galaxies at $z \sim$ 4 ($m^*$ = 23.31 $\pm$ 0.08) \citep{bou15}. The median magnitude in F105W for our sample is 24.78. 
 
The rest-frame band of F105W at the median redshift of Mg~{\sc ii} ($z$ = 3.743) is $U$ band. The LBGs UV-continuum slope $\gamma$ ($f_{\lambda}$ = $\lambda^{\gamma}$) measured by \citet{bouwens12} at $z\sim4$ is around $-$2, therefore we have $u-b\sim$ 0. If the galaxy is a passive galaxy, it is unable to be detected with present images. We then obtained $L_B/L_B^{*}=$ 0.25, where $L_B$ and $L_B^*$ are the $B$ band luminosity of our galaxy candidates and $L^*$ galaxies, respectively. The result is consistent with the estimates of the \Mgtwo associated galaxy luminosity in M13. 

Because we only have a single photometric band measurement, we do not know their redshifts or whether they are associated with the detected absorption systems. We search the archival images of the Dark Energy Camera Legacy Survey (DECaLS) \citep{dey19} and find that four quasar fields are covered by DECaLS (J0002+2550, J0842+1218, J1207+0630, J1250+3130). None of the above galaxies are detected in the $grz$ bands. The 2$\sigma$ detection limit in the $g$ band (the deepest band) is roughly 25.5 mag (see also Table \ref{table:hst}). The red $g$--F105W color implies that these galaxies are likely at high redshift. We further estimate the surface density of $z>$ 2.5 galaxies brighter than 25.5 mag in a few \hst~ fields \citep{bou15} and find that the expected number of random galaxies in a 5$\arcsec$ circular area is about 0.1. This is significantly lower than our density of $>1$, suggesting that the galaxies detected in F105W above are likely associated with the strong absorbers. We can see that within 50 kpc of each absorber, we detected maximum two galaxy candidates. If multiple galaxies are associated with strong \Mgtwo systems at high redshift, we need higher resolution spectra to disentangle the absorbing gas kinematic structure and deeper images to search for galaxy candidates nearby. 

\subsection{Gas halo size and galaxy impact parameter}

We have limited our search of galaxy candidates within a 5$\arcsec$ (42.36 kpc at $z$ = 2.2) radius, and we detected at least one candidate for each absorber within an impact parameter $D = 50$ kpc. This median distance of 23.31 kpc is smaller than the median distance, $<D>$ = 48.7 kpc found in the local Universe \citep{sch12,nie13b,nie13a,nie18}, suggesting that strong \Mgtwo absorbers likely have smaller impact parameters at higher redshift.
We can also calculate the absorbing halo gas size from the measured $dN/dX$ and $L_B/L_B^*$ using the relation from \citet{kac08}. The comoving line density $(dN/dz)$ can be expressed as the product of the absorber physical cross-section $\sigma$ and volume co-moving number density $n(z)$,
\begin{equation}\label{eq:dndz}
\frac{dN}{dz} = \frac{c}{H_0} \sigma n(z) \frac{dX}{dz},
\end{equation}   
where c/${H_0}$ is the constant of proportionality and
\begin{equation}
    \frac{dX}{dz} = \frac{(1+z)^2}{\sqrt{\Omega_m(1+z)^3+\Omega_{\Lambda}}}.
\end{equation}
The gas cross-section is expressed as $\sigma$ = $\pi R_x^2$, where $R_x$ is the absorbing gas halo size. The volume number density $n(z)$ can be expressed as a function of associated galaxy luminosity:
\begin{equation}
    n(z) = \Phi^*\times\Gamma(x,y),
\end{equation}
where $\Phi^*$ is the number density of $L^*$ galaxies in the galaxy luminosity function. $\Gamma(x,y)$ is an incomplete Gamma function with $x  = 2\beta-\alpha+1$, where $\alpha$ is the faint-end slope of Schechter function and $\beta$ is the factor of a relation between $R_x$ and associated galaxy luminosity \citep{ste95}: $R_x = R_* \times (L/L^*)^{\beta}$. $y$ is the ratio of the detected galaxy minimum luminosity to $L^*$.
Therefore $R_x$ is:
\begin{equation}
 R_x = \sqrt{\frac{dN/dX}{\pi\Phi^*\Gamma(x,y)}} .
 \end{equation}
%\begin{equation}
%\frac{dN}{dz}  = \pi R_x^2 \Phi^* \Gamma(x,y) \frac{c}{H_0} %\frac{(1+z)^2}{\sqrt{\Omega_m(1+z)^3+\Omega_{\Lambda}}},
%\end{equation}
%With an empirical relation between absorbing cloud size $R_x$ and $B$-band luminosity of associated galaxies at the local Universe was reported in \citet{chen10}, {\bf which analysed 47 Mg~{\sc ii} associated galaxies at $z<$ 0.5 and with 0.1 $<$\wmgone $<$ 2.34 \AA}: $R_\textrm{{Mg~{\sc ii}}}$ = 75 $\times(L_B/L_B^*)^{0.35\pm0.03}$. {\bf This relation fits for both star-forming and passive galaxies.} We assume that the relation does not evolve at high redshift, 
We take $\alpha  = -1.64\pm0.04$ from \citet{bou15}, $\beta = 0.35$,  our measured $y = L_B/L_B^* =$ 0.25 and $dN/dz$ at $<z>$ = 3.5, $\Phi^*_B$ = 1.97 $^{+0.34}_{-0.28}\times 10^{-3} $ Mpc $^{-3}$ \citep{bou15} to calcualte the $R_x$. The $\beta$ value is from \citet{chen10}, which analysed 47 Mg~{\sc ii} associated galaxies at $z<$ 0.5 and with 0.1 $<$\wmgone $<$ 2.34 \AA. We assume the slope does not change at higher redshift. The $R_x$ is then estimated as follows: 
\begin{equation}\label{eq:Rx}
	    R_x \textrm{(kpc)} = \left\{
	    \begin{array}{cl}
		37, & \beta = 0.35, y = 0.05 ; \\
		8,&\beta = 0.35, y = 0.25.
		\end{array}
		\right.
\end{equation}
%12,& \beta = 0.35, y = 0.1.\\
With our measured $L_B/L_B^* =$ 0.25, $R_x$ is smaller than the possible $D$. %If the $R_x$--$L$ has not evolution has no evolution at high redshift and fainter galaxies ($L_B/L_B^*$ = 0.05) associated with Mg~{\sc ii} absorberss, the $R_x$ is comparable or slightly larger than $D$. 
This is based on the assumption that the covering fraction $f_c$ of the gas is unity. Covering fraction of the absorbing gas is defined as the ratio of absorbers associated galaxies and all galaxies at the same redshift bin. \citet{lan20} found that covering fraction of strong Mg~{\sc ii} systems evolves with redshift at 0.4 $<z<$ 1.3, similarly to the evolution of star-formation rate of galaxies. In the study of \citet{chen10}, the Mg~{\sc ii}-absorbing halo gas covering fraction is 70\% for \wmgone $>$ 0.3 \AA. \citet{nie18} studied 74 galaxies at 0.113 $<z<$ 0.888 with $\langle$\wmgone$\rangle$ = 0.65 \AA, and found $f_c$ = 0.68 for isolated galaxies. Since we have 1-2 galaxy candidate within 5$\arcsec$ of the absorber, we adopt the $f_c$ = 0.68. The covering fraction corrected size $R_*$ is 9.23 kpc ($R_x^2$ = $f_cR_*^2$), which is still smaller than the $<D>$ = 23.31 kpc.

%{\bf We discuss above with the presumptions that our associated galaxy candidates are isolated (maximum two) star-forming galaxies. If they are associated with e.g. luminous red galaxies, the covering fraction is down to e.g. 10\% \citep{lan14}. Thus, then $R_*\sim D$ is possible.}

In summary, we searched for associated galaxies around our strong Mg~{\sc ii} absorbers at $z$ = 3--5.1 within 50 kpc and found 1--2 candidates for each absorber. The galaxy candidates are brighter than 25.5 mag and have a median magnitude of 24.78 in F105W  band. Theses candidates have a median impact paramter $<D>$ = 23.31 kpc, which is smaller than that at $z<$  1. If we assume that the $R_x$--$L$ slope and $f_c$ for strong Mg~{\sc ii} absorbers at $z$ = 3--5.1 are similar to that at $z<$ 1, with a fixed associated galaxy luminosity $L = 0.25 L*$ and a covering fraction of $f_c$ = 0.68, the $f_c$-corrected absorbing halo gas $R_*$ is smaller than the $<D>$. In another word, within 50 kpc, high redshift strong Mg~{\sc ii} absorbers tend to have a more disturbed environment but smaller halo size than that at $z<1$.

%\subsection{Possibility of passive galaxies}
% However, early-type galaxies also host Mg~{\sc ii} absorbers \citep{chen10b,gau10,gau11}. Though weaker systems are presented in \citet{gau11}, which studies 37 Luminous Red Galaxies (LRG) and found eight has associated Mg~{\sc ii} absorbers with \wmgone $>$ 0.3 \AA and velocity separation smaller than 350 km s$^{-1}$. \citet{werk13} analysed 44 LRGs around with Mg~{\sc ii} systems and found the $f_c$ out to 50 kpc for \wmgone $>1$ \AA is smaller than 30\%. \citet{lan14} analsysed the strong Mg~{\sc ii} systems associated with LRGs, the results indicates that the covering fraction for \wmgone $>$ 1\AA systems is $\sim$ 0.1. Recall our 

%The $R_x$ is correlated with galaxy luminosity over $L^*$ \citep{ste95},
%\begin{equation}
%R_x = R_* \times (L/L^*)^{\beta}.
%\end{equation}

 %With the assumption that \Mgtwo absorbers are associated with Lyman Break Galaxies (LBGs) at high redshift, the $dN/dX$ decline %of strong \Mgtwo absorbers at $z>$ 3 can be explained by the $n(z)$ decline of LBGs  (see \citet{bou15}) and/or the decline of %the \Mgtwo gas cross-section. We calculate \Mgtwo$\sigma$ using a relation between the expected galaxy number and the $dN/dX$ %of strong \Mgtwo absorbers. We use the LBG luminosity function in \citet{bou15} at $<z>$ = 3.8, 4.9, 5.9, and we assume %$L_{B_{min}}$ = 0.25 $L^*$. The resultant cross sections are 0.02, 0.007 and 0.003 Mpc$^2$ e at $<z>$ = 3.8, 4.9, 5.9 %respectively. This suggests that the $dN/dX$ decline of \Mgtwo  is affected by both the gas cross-section and associated LBG %volume density.

\subsection{Individual Systems}\label{sec:individual}
In this subsection, we present a few individual systems with peculiar absorption features or having images in other bands.

{\it J0050+3445.}
We detected a \Mgtwo absorber at $z$ = 3.435 with \wmgone =3.44 $\pm$ 0.88\AA. There are two galaxy candidates within 5$\arcsec$ from the quasar, labeled as 1 and 2 in Figure \ref{fig:J0050_J2310_img}. We measure the $g$ and $r$ band magnitudes of the two objects using archived Canada France Hawaii Telescope (CFHT) MegaPrime images. The magnitudes of galaxy 1 are $g$ = 25.44 $\pm$ 0.10 and $i$ =25.64 $\pm$ 0.11. The magnitudes of galaxy 2 are $g$ = 25.43 $\pm$ 0.10 and $i$ =24.89 $\pm$ 0.13. The Ly$\alpha$ emission line at $z$ = 3.435 is redshifted to 5391 \AA\ ($g$ band), so galaxy 2 with $g-i$ = 0.54 is more likely to be the absorber.

{\it J2310+1855}. 
We detected strong \Mgtwo and \Fetwo at \z = 4.244. In Figure \ref{fig:J0050_J2310_img} there are two galaxies within 5$\arcsec$ (25 kpc) from the quasar. The magnitudes of galaxy 1 are $r$ = 25.55 $\pm$ 0.14, $z$ =24.34 $\pm$ 0.25, and $F105W$ = 24.26 $\pm$ 0.05. The magnitudes of galaxy 2 are $r$ = 26.18 $\pm$ 0.25 and $F105W$ = 25.10 $\pm$ 0.11. It is not detected in $z$. Based on their colors, galaxy 1 is more likely to be the absorber host at $z = 4.244$. In addition, there is a bright object next to galaxy 1. But its $m_z$ = 22.15 is much brighter than $L^*$  at $z\sim$ 4, so we did not consider it.

%{\it J0927+2001}. There are two galaxy candidates within 50 kpc in the {\it HST} images. However, there is no \Mgtwo and \Fetwo absorption detected.

\begin{figure}
  \resizebox{\hsize}{!}{\includegraphics{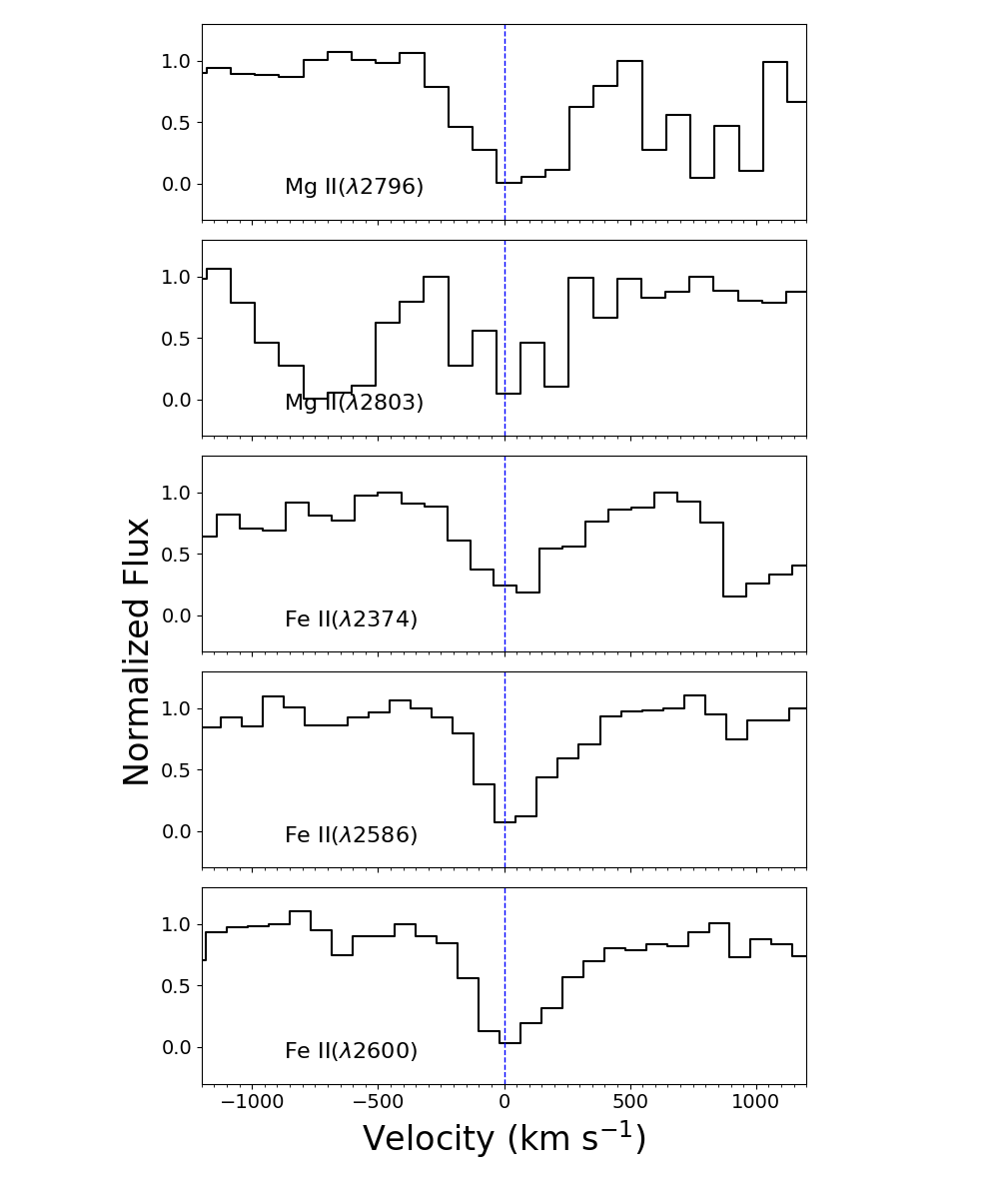}}
  \caption{\small{ Velocity profile of J1148+0702 at $z$ = 4.369. This system has \Mgtwo absorption (\wmgone = 4.23 $\pm$ 0.52 \AA) and the strongest \Fetwo absorption (\wfeone = 3.61 $\pm$ 0.54 \AA) in the whole sample. However, there is no image data for this quasar in HST, CFHT, DECaLS and Pan-STARRS archives. 
}}\label{fig:J1148}
\end{figure}

{\it J1148+0702}. We detected \Mgtwo and \Fetwo at \z = 4.369  and $z$  = 3.495, the latter one has extremely large velocity spread for a Mg~{\sc ii} doublet. This one at $z$ = 4.369 has a very strong \Mgtwo (\wmgone $>$ 4 \AA~) absorber with the strongest \Fetwo ($W_r (\lambda 2600)$ = 4.45 $\pm$ 0.82 \AA~) absorption in our sample. Both \Mgtwo and \Fetwo lines are strongly saturated. As discussed in \citet{jos17}, strong \Mgtwo and \Fetwo in the same system may indicate star formation nearby. We do not have {\it HST} images for this quasar. The \Mgtwo and \Fetwo absorption profiles are presented in Figure \ref{fig:J1148}.

%{\it J1429+5447}. In the {\it HST} image (see Figure \ref{fig:hst}), there are two extended sources detected within a radius of 5$\arcsec$ of the quasar , however, we did not detect strong \Mgtwo and \Fetwo systems with our spectrum.

%{\it J1602+4228}. There is a galaxy 0.85$\arcsec$ away from the quasar center (see Figure \ref{fig:hst}). However, we did not detect strong \Mgtwo or \Fetwo absorption possibly associated with this galaxy. 

%---------------------------------------------------------------------------------------------------------------------------------------
\section{Conclusion}

We have analyzed the near-IR spectra of 31 luminous quasars at $z>5.7$, selected from a sample of 50 quasars observed by {\it Gemini} GNIRS. We identified 32 \Mgtwo and Fe~{\sc ii} absorbers with \Mgtwo $W_{r}$(2796) $>1.0$ \AA\ at $2.2<z<6.0$. We calculated the line density $dN/dz$ and comoving line density $dN/dX$ of the strong \Mgtwo absorbers and found that they decrease towards higher redshift at $z>3$. This can be described by the relation $dN/dz$ = (1.882$\pm$3.252)$\times(1+z)^{-0.952\pm1.108}$. The trend is consistent with previous results, and follows the evolution of the cosmic star formation rate, implying the correlation between strong \Mgtwo absorbers with the star formation of galaxies at high redshift.

We found that 15/32 of our \Mgtwo systems have large velocity widths with $\Delta v >$ 300 \kms, which is much larger than those detected in DLA systems with similar equivalent widths at 2$<z<$ 4 and \Mgtwo systems at $z <$ 2. Such large velocity widths are also seen in a sample of neutral-carbon selected \Mgtwo systems at 1.5 $<z<$ 2.7. This potentially implies that strong \Mgtwo systems at high redshift are influenced by galactic superwinds and/or interaction within galaxy groups. Also, our Mg~{\sc ii} systems exhibit slightly less-saturation in terms of the equivalent width ratio of Fe~{\sc ii} and Mg~{\sc ii} lines (\wfeone/\wmgone). For DLA systems, \wfeone/\wmgone $\sim0.5$. This ratio is roughly between 0.25 and 1.75 in our sample. Our \Mgtwo absorbers are possibly less saturated than DLA-\Mgtwo at 2 $<z< $ 4 and those at $z<$ 2.3 with similar equivalent widths. This is potentially caused by the interaction of more sub-components of our strong Mg~{\sc ii} systems.

We have used several {\it HST} images (together with archival DECaLS and CFHT images) to identify potential absorber galaxies within 50 kpc from quasars. For \Mgtwo systems have $\Delta v>$ 300 \kms or \wmgone $>$ 1.5 \AA, there are  1-2 galaxy candidates within the 5$\arcsec$ radius. The median F105W-band magnitudes is 24.83 mag, which is fainter than the $L^*$ galaxy luminosity at $z\sim$ 4. If the Mg~{\sc ii}-absorbing halo gas and associated galaxy luminosity relation at $z =$ 3--5 is similar to that at $z<$ 1, the \Mgtwo absorbing gas size $R_x$ is smaller than $D$.  
%With our detected galaxy candidate luminosity and the associated strong \Mgtwo comoving line density at similar redshift, we suggest that the cross-section of strong \Mgtwo associated gas decreases with redshift at $z>$ 3.

\acknowledgments
We thank the very constructive comments and suggestions from the anonymous referee. We thank Patrick Petitjean for useful comments and Robert A. Simcoe for discussion on FIRE data. We acknowledge support from the National Science Foundation of China (11533001, 11721303, 11890693, 11991052), the National Key R\&D Program of China (2016YFA0400702, 2016YFA0400703), and the Chinese Academy of Sciences (CAS) through a China-Chile Joint Research Fund \#1503 administered by the CAS South America Center for Astronomy in Santiago, Chile. Y.S. acknowledges support from an Alfred P. Sloan Research Fellowship and National Science Foundation grant No. AST-1715579. M.V. gratefully acknowledges financial support from the Independent Research Fund Denmark via grant number DFF 8021-00130. We acknowledge the public data from the Dark Energy Camera Legacy Survey (DECaLS), the Beijing-Arizona Sky Survey (BASS), and the Mayall $z$-band Legacy Survey (MzLS). We thank observations obtained with MegaPrime/MegaCam, a joint project of CFHT and CEA/DAPNIA, at the Canada-France-Hawaii Telescope (CFHT) which is operated by the National Research Council (NRC) of Canada, the Institut National des Science de l'Univers of the Centre National de la Recherche Scientifique (CNRS) of France, and the University of Hawaii.

\facilities{Gemini(GNIRS)}

\clearpage



%\bibliographystyle{mnras}


\newpage
\bibliographystyle{aasjournal}
\bibliography{./gnir}
%%%%%%%%%%%%%%%%%%%%%%%%%%%%%%%%%%%%%%%%%%%%%%%%%%

%%%%%%%%%%%%%%%%% APPENDICES %%%%%%%%%%%%%%%%%%%%%
% Don't change these lines
%\bsp	% typesetting comment
\end{document}